\documentclass[a4paper, 12pt, oneside]{article}
\usepackage{amsmath}
\usepackage{amssymb}
\usepackage{enumerate}
\usepackage{tikz}
\usetikzlibrary{matrix}

\newcounter{thm} 

\newtheorem{Thm}[thm]{Theorem}

\newtheorem{Lem}[thm]{Lemma}
\newtheorem{Prop}[thm]{Proposition}
\newtheorem{Cor}[thm]{Corollary}

\def \Z {\mathbb Z}

\def \R {\mathbb R}

\def \g {\mathfrak g}

\begin{document}
\title{Connecting the Kontsevich-Witten and Hodge tau-functions by the $\widehat{GL(\infty)}$ operators}
\author{Xiaobo Liu \thanks{Research of the first author was partially supported by  NSFC research fund 11431001,NSFC Tianyuan special funds 11326023 \& 11426226, SRFDP grant 20120001110051.}, \hspace{6pt} Gehao Wang}

\date{}

\maketitle

\abstract
In this paper, we present an explicit formula that connects the Kontsevich-Witten tau-function and the Hodge tau-function by differential operators belonging to the $\widehat{GL(\infty)}$ group. Indeed, we show that the two tau-functions can be connected using Virasoro operators. This proves a conjecture posted by Alexandrov in \cite{A}.

\section{Introduction}

The tau-functions have been playing a very important role in the study of integrable systems. One of the known facts is that the set of all tau-functions of the KP hierarchy forms an orbit under the action of the group $\widehat{GL(\infty)}$. This group is associated to the infinite dimensional Lie algebra $\widehat{\mathfrak{gl}(\infty)}$, which is a central extension of the $\mathfrak{gl}(\infty)$ algebra. Many well-known tau-functions arose in algebraic geometry, such as the Kontsevich-Witten tau-function, Hodge tau-function and Hurwitz tau-function.
Since the group $\widehat{GL(\infty)}$ preserves the KP integrability, relations between such tau-functions using operators of this type are particularly interesting.
 For example, Kazarian proved in \cite{MK}  that the Hodge tau-function $\exp(F_H(u,q))$ is indeed a tau-function for the KP hierarchy, where $F_H(u,q)$ is obtained from the Hodge series $F_H(u,t)$ after a change of variables derived in \cite{MK}.  The main idea of his proof was to show that $\exp(F_H(u,q))$ is connected to the Hurwitz tau-function $\exp(H)$ by a $\widehat{GL(\infty)}$ operator, where $H$ is the generating function of simple Hurwitz numbers and it was known that $\exp(H)$ is a tau-function of the KP hierarchy (cf. \cite{AO},\cite{KL}).

The Kontsevich-Witten tau-function, denoted by $\exp(F_K(q))$, satisfies the KdV hierarchy (cf. \cite{K}). The function $F_K(q)$ is obtained from the generating function $F_K(t)$ for the intersection numbers of $\psi$-classes after the variable change $t_i=(2i-1)!!q_{2i+1}$. Using Mumford's theorem in \cite{Mu}, one can transform $F_K(t)$ into the Hodge series $F_H(u,t)$ using a differential operator $W$ described by equation (\ref{W}), namely,
\begin{equation}\label{1}
\exp(F_H(u,t))=e^W\cdot \exp(F_K(t)).
\end{equation}
However, $W$ does not belong to the $\widehat{\mathfrak{gl}(\infty)}$ algebra. Recently, Alexandrov posted a conjecture in \cite{A} stating that there exists a $\widehat{GL(\infty)}$ operator connecting the Kontsevich-Witten tau-function $\exp(F_K(q))$ and Hurwitz tau-function $\exp(H)$. This operator consists of two parts, where the first part connecting $\exp(H)$ and $\exp(F_H(u,q))$ is already know in \cite{MK}. Note that the tau-function $\exp(H)$ has a description of the cut-and-join operator \cite{GJ1}, which belongs to the $\widehat{GL(\infty)}$ group. Hence, if this conjecture is proved, not only will it allow us to have a $\widehat{GL(\infty)}$ operator description of the Kontsevich-Witten tau-function, but also provides a possible way to derive the Virasoro constraints for the Hurwitz tau-function. Later, Alexandrov refined this conjecture as Conjecture 2.1 in \cite{AE} stating that there exists a $\widehat{GL(\infty)}$ operator $\widehat{G}_{+}$ formed by Virasoro operators, such that
$$\widehat{G}_{+}\cdot\exp(F_H(u,q))=\exp(F_K(q)).$$
He also verified this conjecture up-to an unknown factor, which is a Taylor series in variable $u$, and hence confirmed a weaker version of this conjecture (see Section 4).

In this paper, we give a complete proof of Alexandrov's conjecture in \cite{A}. Our method is totally different from the method used in \cite{AE}. In our proof, we can see a clear relation between actions of the differential operator $W$ and the Virasoro operators. We will also give a more explicit way for computing coefficients of the Virasoro operators. We first present an explicit $\widehat{GL(\infty)}$ operator that connects the two tau-functions $\exp(F_H(u,q))$ and $\exp(F_K(q))$:
\begin{Thm}\label{main}
The relation between the two tau-functions $\exp(F_H(u,q))$ and $\exp(F_K(q))$ can be written as the following:
\begin{equation}\label{maineq}
\exp(F_H(u,q))=\exp{(\sum_{m>0} a_mu^mL_m)}\exp(P)\cdot\exp{(F_K(q))},
\end{equation}
where $L_m,m\geq 1,$ is the {\bf Virasoro operator}
$$L_m=\sum_{k>0,k+m>0} (k+m)q_k\frac{\partial}{\partial q_{k+m}}+\frac{1}{2}\sum_{a+b=m} ab\frac{\partial^2}{\partial q_a\partial q_b}.$$
The coefficients $\{a_m\}$ can be computed from the equation
\begin{align} \label{eqn:a}
\exp(\sum_{m>0}a_mz^{1-m}\frac{\partial}{\partial z})\cdot z&=\left(-2\log(1-\frac{1}{1+z})-\frac{2}{1+z}  \right)^{-\frac{1}{2}}.
% \\
% &=z+\frac{2}{3}-\frac{1}{12}z^{-1}+\dots .
\end{align}
 And $P$ is a first order differential operator of the form
$$P=-\sum_{k=1}^{\infty} b_{2k+1}u^{2k}\frac{\partial}{\partial q_{2k+3}},$$
where the numbers $\{b_{2k+1}\}$ are uniquely determined by the recursion relation,
\begin{equation}\label{re}
(n+1)b_n=b_{n-1}-\sum_{k=2}^{n-1}kb_kb_{n+1-k}
\end{equation}
with $b_1=1, b_2=1/3$.
\end{Thm}

The coefficients $\{a_m\}$ are first determined by Lemma \ref{P5} below. In Section 3, we will prove this main result. The method is to decompose the two operators $\exp(W)$ and $\exp{(\sum_{m>0} a_mu^mL_m)}$ into several factors using the Zassenhaus formula, then study the action of these factors and compare them. One of the difficulties in the proof lies on the fact that $F_{K}(q)$ only involves odd variables $q_{2i+1}$ while $F_{H}(u, q)$ involves all variables $q_i$. This causes the problem that the decompositions of the above two operators do not exactly match with each other. We will show that the differences of these operators do not matter when acting on
the function $\exp{(F_K(q))}$. For this purpose, we need to implement an integral transform on functions using Gaussian integrals. We also need to use properties of the gamma and beta functions. More details can be found in Section 3.3.

We can also use an explicit function $h(z)$ defined in Corollary \ref{hz} to determine the coefficients $\{a_m\}$. The power series $h$ and the function on the right hand side of equation (\ref{eqn:a}) are both  related to the Lambert W function, about which we will explain more in Section 2.3. After a brief introduction in Section 2.2, we can see that the Virasoro operators and $P$ all belong to the infinite dimensional Lie algebra $\widehat{\mathfrak{gl}(\infty)}$ acting on the space of KP solutions. Therefore the KP integrability is preserved in our expression. Note that the operator $\exp(P)$ performs a shift of arguments on a series. The extra parameter $u$ in $P$ guarantees that $\exp(P)\cdot \exp(F_K(q))$ is a well-defined formal series. Furthermore, due to the fact that the Kontsevich-Witten tau-function satisfies the Virasoro constraints,
an application of Theorem~\ref{main} is the following
\begin{Cor}\label{em}
There exists a sequence of numbers $\{e_m\}$, such that
$$\exp(F_H(u,q))=\exp{(\sum_{m>0} e_mu^mL_m)}\cdot\exp(F_K(q)).$$
The coefficients $\{e_m\}$ are determined by equation (\ref{emtheta}) below.
\end{Cor}
This gives a complete proof of Alexandrov's conjecture in \cite{A}. We have computed the first few terms of $\{e_m\}$ by hand, and they coincide with Alexandrov's data. We will explain this in Section 4.
Moreover, since the Kontsevich-Witten tau-function satisfies the KdV hierarchy, Theorem \ref{main} gives an alternative proof for Kazarian's theorem that $\exp(F_H(u,q))$ is a tau-function of the KP hierarchy.
This proof does not need the theory of Hurwitz numbers and ELSV formula.

\section*{Acknowledgement}
The authors would like to thank Yizhi Huang for helpful discussions. They would also like to thank the referee for suggestions in improving the presentation of this paper.

\section{Preliminaries}
\subsection{Kontsevich-Witten and Hodge series}
Let $\overline{M}_{g,n}$ be the moduli space of complex stable curves of genus $g$ with $n$ marked points, and $\psi_i$ be the first Chern class of the cotangent space over $\overline{M}_{g,n}$ at the $i$th marked point. The intersections of the $\psi$-classes are evaluated by the integral:
$$<\tau_{d_1}\dots \tau_{d_n}>=\int_{\overline{M}_{g,n}}\psi_1^{d_1}\dots \psi_n^{d_n}.$$
The Kontsevich-Witten generating function in variables $t_k$ is defined as
$$F_K(t)=\sum <\tau_0^{k_0}\tau_1^{k_1}\dots>\frac{t_0^{k_0}}{k_0!}\frac{t_1^{k_1}}{k_1!}\dots.$$
Let $\lambda_j$ be the $j$th Chern class of the Hodge bundle over $\overline{M}_{g,n}$ whose fibers over each curve is the space of holomorphic one-forms on that curve. The Hodge integrals are the intersection numbers of the form
$$<\lambda_j\tau_{d_1}\dots \tau_{d_n}>=\int_{\overline{M}_{g,n}}\lambda_j\psi_1^{d_1}\dots \psi_n^{d_n}.$$
They are defined to be zero when the numbers $j$ and $d_i$ do not satisfy the condition
\begin{equation*}
j+\sum_{i=1}^n d_i=dim(\overline{M}_{g,n})=3g-3+n.
\end{equation*}

Hodge series is defined as
$$F_H(u,t)=\sum (-1)^j <\lambda_j \tau_0^{k_0}\tau_1^{k_1}\dots>u^{2j} \frac{t_0^{k_0}}{k_0!}\frac{t_1^{k_1}}{k_1!}\dots$$
where $u$ is the parameter marking the $\lambda$-class. In fact, $F_H(0,t)=F_K(t)$. As we have mentioned before, the two functions $F_K(t)$ and $F_H(u,t)$ are related by operator $\exp(W)$ as in equation (\ref{1}), (cf.\cite{FP},\cite{AG1}), where
\begin{equation}\label{W}
W=-\sum_{k\geq 1} \frac{B_{2k}u^{2(2k-1)}}{2k(2k-1)}(\frac{\partial}{\partial t_{2k}}-\sum_{i\geq 0} t_i\frac{\partial}{\partial t_{i+2k-1}}+\frac{1}{2}\sum_{i+j=2k-2} (-1)^i\frac{\partial^2}{\partial t_i\partial t_j}).
\end{equation}
Here $B_{2k}$ is the Bernoulli numbers defined by:
$$\frac{t}{e^t-1}=\sum_{m=0}^{\infty} B_m\frac{t^m}{m!}.$$
Note that $W$ does not belong to the $\widehat{\mathfrak{gl}(\infty)}$ algebra.

Now let
$$F_K(q)=\left.F_K(t)\right\vert_{t_k=(2k-1)!!q_{2k+1}}.$$
It is well-known that $\exp{(F_K(q))}$ is a tau-function for the KdV hierarchy \cite{K}. And $\exp({F_H(u,t)})$ is a tau-function for the KP hierarchy after the a change of variables derived in \cite{MK}. It can be described in the following way. Let
\begin{equation}\label{D}
\widehat{D}=(u+z)^2z\frac{\partial}{\partial z}.
\end{equation}
Consider the following sequence of polynomials $\phi_k(u,z)$:
$$\phi_0(u,z)=z,\quad\quad \phi_{k}(u,z)=\widehat{D}^{k}z=\sum_{j=1}^{2k+1} \alpha_{j}^{(k)}u^{2k+1-j}z^j,$$
for some constants $\alpha_{j}^{(k)}$, where $\alpha_{2k+1}^{(k)}=(2k-1)!!$. Let $\widetilde{\phi_k}(u,q)$ be the polynomial in variables $q_i$ and $u$ obtained by replacing $z^m$ with $q_m$ in $\phi_k(u,z)$. Then the variable change is in the form
$$t_k=\widetilde{\phi_k}(u,q)=\sum_{j=1}^{2k+1} \alpha_{j}^{(k)} u^{2k+1-j}q_j,$$
For example,
\begin{align*}
t_0 & = q_1\\
t_1 & = u^2q_1+2uq_2+q_3\\
t_2 & = u^4q_1+6u^3q_2+12u^2q_3+10uq_4+3q_5\\
t_3 & = u^6q_1+14u^5q_2+61u^4q_3+124u^3q_4+131u^2q_5+70uq_6+15q_7\\
\dots .
\end{align*}
Note that, by definition of the double factorial, $0!!=(-1)!!=1$. Let
$$F_H(u,q)=\left.F_H(u,t)\right\vert_{t_k=\widetilde{\phi_k}(u,q)}.$$
If we set $u=0$, the polynomial $\widetilde{\phi_k}(u,q)$ only has the term $(2k-1)!!q_{2k+1}$, and $\exp(F_H(0,q))$ becomes the Kontsevich-Witten tau-function $\exp(F_K(q))$.

\subsection{The transformation group $\widehat{GL(\infty)}$}
The space of tau-functions for the KP hierarchy forms a group orbit \cite{DJKM}. We denote this group by $\widehat{GL(\infty)}$, and it is defined from the infinite dimensional Lie algebra $\widehat{\mathfrak{gl}(\infty)}$ via the exponential map. Note that, strictly speaking, the ``group elements'' in $\widehat{GL(\infty)}$ constructed in this way do not belong to a well-defined group, because the products of these elements might be divergent. Hence when we want to use a ``group element'' formed by a product of other such elements, we need to check whether this product is well-defined.

Here, we introduce some sample operators that belong to the Lie algebra $\widehat{\mathfrak{gl}(\infty)}$ and will be used in our results. We refer the readers to \cite{MJE} or other related articles for more details. Suppose the tau-functions are in variables $q_k$. The operators $q_k$ (i.e. multiplication by $q_k$) and $\partial/\partial q_k$ belong to $\widehat{\mathfrak{gl}(\infty)}$. It follows that adding a linear function to the solution or a shift of arguments preserve the solutions of the KP hierarchy. The Virasoro operators $\{L_m\}$ that generate the Virasoro algebra with central charge $c=1$ also belong to this Lie algebra. For $m\in\Z$, setting $q_j=0$ and $\partial/\partial q_j=0$ for $j<1$, we can write $L_m$ as
$$L_m=\sum_{k>0,k+m>0} (k+m)q_k\frac{\partial}{\partial q_{k+m}}+\frac{1}{2}\sum_{a+b=m} ab\frac{\partial^2}{\partial q_a\partial q_b}+\frac{1}{2}\sum_{a+b=-m}q_aq_b,$$
and they satisfy the commutator relation
$$[L_m,L_n]=(m-n)L_{m+n}+\frac{1}{12}\delta_{m,-n}(m^3-m).$$
In particular, the degree operator $L_0=\sum kq_k\partial/\partial q_k$. In Section 3, we will discuss how we deal with the exponential of Virasoro operators when they act on functions.

\subsection{Gamma function and beta function}
The {\bf gamma function} $\Gamma(z)$ is an extension of the factorial function. If $z$ is a positive integer $n$, then
$$\Gamma(n+1)=n!.$$
And when $z$ is a complex number with positive real part, then the following equation holds
$$\Gamma(z+1)=z\Gamma(z).$$
It can be defined by a definite integral, known as the {\bf Euler integral of the second kind},
\begin{equation}\label{euler2nd}
\Gamma(z) = \int_0^\infty  x^{z-1} e^{-x}\mathrm{d}x.
\end{equation}
This definition is only valid for complex number $z$ with positive real part. The function $\Gamma(z)$ can be extended to a meromorphic function on the whole complex plane with simple poles at non-positive integers.
%The famous {\bf Euler's reflection formula}
%$$\Gamma(z)\Gamma(1-z) = \frac{\pi}{\sin{(\pi z)}},$$
%describes a relationship between $\Gamma(1-z)$ and $\Gamma(z)$ when $z$ is not an integer.
Furthermore, using {\bf Stirling's approximation}, one can obtain an asymptotic expansion for $\Gamma(z)$ with the value of $|z|$ large enough:
\begin{equation}\label{gamma}
\Gamma(z)\sim z^{z-\frac{1}{2}}e^{-z}\sqrt{2\pi}\sum_{i=0}^{\infty}C_iz^{-i},
\end{equation}
with $C_0=1$. The coefficients $\{C_i\}$ play an important role here in this paper. In fact,
\begin{equation}\label{Cb}
C_i=(2i+1)!!b_{2i+1},
\end{equation}
where the coefficients $\{b_{2i+1}\}$ appear in the relation (\ref{re}) (cf. \cite{MM}). If we take the logarithm on $\Gamma(z)$, the coefficients $B_{2k}$ appearing in the expansion are simply the Bernoulli numbers:
$$\log (\Gamma (z)) \sim \left(z-\tfrac{1}{2}\right)\log(z) -z + \tfrac{1}{2}\log(2 \pi) + \mathfrak{B}(z),$$
where
\begin{equation*}
\mathfrak{B}(z)=\sum_{k=1}^\infty \frac{B_{2k}}{2k(2k-1)}z^{-2k+1}.
\end{equation*}
This implies
\begin{equation}\label{B2}
e^{\mathfrak{B}(z)} =\sum_{i=0}^{\infty}C_iz^{-i}.
\end{equation}
If we expand the left hand side of the above equation, we obtain the general formula as
$$e^{\mathfrak{B}(z)} = 1+\sum_{m>0} \frac{1}{m!}\sum_{k_1,\dots k_m\geq 1} \prod_{j=1}^m \frac{B_{2k_j}}{2k_j(2k_j-1)} z^{m-\sum_{j=1}^m 2k_j}.$$
Hence
\begin{equation}\label{B3}
C_i=\sum_{m=1}^{i} \frac{1}{m!} \sum_{\substack{\sum_{j=1}^m 2k_j=i+m \\ k_1,\dots k_m\geq 1} }\prod_{j=1}^m \frac{B_{2k_j}}{2k_j(2k_j-1)} .
\end{equation}
The fact that $\mathfrak{B}(z)$ is an odd function immediately implies the following relation on the coefficients $\{C_i\}, i\geq 0$.
\begin{Lem}\label{C}
For all $k\geq 1$, we have
$$\sum_{i=0}^{k} (-1)^iC_iC_{k-i}=0.$$
\end{Lem}
Furthermore, for $i\geq 1$, $C_i$ has some nice combinatorial interpretations. It can be explicitly given by
$$C_i=\sum_{k=1}^{2i} (-1)^k \frac{d_3(2i+2k,k)}{2^{i+k}(i+k)!},$$
where $d_3(n,k)$ is the number of permutations on $n$ elements consisting of $k$ permutation cycles with length greater than $2$, (cf. \cite{LC}).

The following results from \cite{MM} gives us an equation that packs all the coefficients $\{b_i\}$ into its solution. We first observe that the function $ve^{1-v}$ is increasing for $v\in [0,1]$, and decreasing for $v\in [1,\infty)$. Now, consider the equation
\begin{equation}\label{vw}
ve^{1-v}=e^{-\frac{1}{2}x^2}.
\end{equation}
The above equation defines a one-to-one relation between $v(x)$ in $[1, \infty)$ and $x$ in $[0,\infty)$. In this case, $v$ increases with $x$, and we can write $v$ as a power series of $x$ in the form
\begin{equation}\label{v}
v=1+\sum_{i=1}^{\infty}b_ix^i.
\end{equation}
The existence of the power series $v(x)$ is guaranteed by the Lagrange inversion theorem \cite{GJ}. Then, differentiate the equation by $x$ on both sides, we have
\begin{equation}\label{vwrelation}
v^{'}(v-1)=xv.
\end{equation}
This gives us exactly the relation (\ref{re}) on the coefficients $\{b_i\}$ with $b_1^2=1$. Since $v^{'}\geq 0$, we have $b_1=1$. The choice of $-1$ for $b_1$ leads to another solution $w(x)$ for equation (\ref{vw}) with $w \in (0,1], x\in [0,\infty)$, and $w$ is decreasing with respect to $x$. In this paper, we always set $b_1=1$. Then $w(x)$ is written as
\begin{equation}\label{w}
w=1+\sum_{i=1}^{\infty} (-1)^ib_ix^i.
\end{equation}
Note that both $v$ and $w$ satisfy the equation (\ref{vwrelation}). Also, it was pointed out in \cite{CJK} that the series $v$ (and $w$) converges for all finite $x$ using the properties of the Lambert W function. In fact, many power series appeared in our context are related to $v$ and $w$.

The {\bf Euler integral of the first kind} defines the {\bf Beta function} $B(x,y)$ for complex numbers $x$ and $y$ with positive real parts:
\begin{equation}\label{euler1st}
B(x,y)=\int_0^1 t^{x-1}(1-t)^{y-1}\mathrm{d}t .
\end{equation}
And it has many other forms including
\begin{equation}\label{betaandgamma}
B(x,y) = \frac{\Gamma(x)\Gamma(y)}{\Gamma(x+y)}.
\end{equation}
All the formulas and results mentioned in this section will be used in our later proofs.

\section{From $F_K(q)$ to $F_H(u,q)$}
Before proving Theorem \ref{main} in this section, we first give an outline for the main ideas of the proof. What we aim to show is that after switched to the variables $q_i$, equation (\ref{1}) implies equation (\ref{maineq}), i.e. the following diagram commutes:
\begin{center}
\begin{tikzpicture}
  \matrix (m) [matrix of math nodes,row sep=3em,column sep=4em,minimum width=2em]
  {
     exp(F_k(t)) & & exp(F_H(u,t)) \\
     exp(F_k(q)) & & exp(F_H(u,q)) \\};
  \path[-stealth]
    (m-1-1) edge  node [left] {{\scriptsize $t_k=(2k-1)!!q_{2k+1}$}} (m-2-1)
            edge node [below] {\scriptsize {$e^W$}} node [above] {{\scriptsize Equation (\ref{1})}} (m-1-3)
    (m-2-1.east|-m-2-3) edge  node [below] {{\scriptsize $e^{\sum_{m>0} a_mu^mL_m}e^P$}} node [above] {{\scriptsize Equation (\ref{maineq})}} (m-2-3)
    (m-1-3) edge node [right] { \scriptsize{ $t_k=\widetilde{\phi_k}(u,q)$ }  } (m-2-3);
\end{tikzpicture}
\end{center}
For this purpose, we want to study the relation between the operators in equation (\ref{maineq}) and $\exp(W)$ after the change of variables. The crucial steps in our proof can be summarized into several propositions, which will be proved later.

Exponential of second order differential operators are very complicated in general. Therefore it is rather difficult to compare operators $\exp(\sum_{m>0} a_mL_m)$ and $\exp(W)$ directly. On the other hand, the exponential of a derivation is much simpler since when acting on a function, it just behaves like a change of variables. Inspired by this fact, we will decompose operators $W$ and $L_m$  into several factors, and first compare the exponential of the derivation part of these operators.  Such comparison will determine the coefficients $a_m$ in Theorem~\ref{main}.

For the operator $W$ defined in equation (\ref{W}). Let
\begin{equation}\label{Ws}
W=\mathfrak{B}_t+\frac{1}{2}Q_0^W+P_0,
\end{equation}
where
\begin{align}\label{Bt}
\mathfrak{B}_t&=\sum_{k=1}^{\infty} \frac{B_{2k}u^{2(2k-1)}}{2k(2k-1)}\sum_{i=0}^{\infty}t_i\frac{\partial}{\partial t_{i+2k-1}},\\
P_0&=-\sum_{k=1}^{\infty} \frac{B_{2k}u^{2(2k-1)}}{2k(2k-1)}\frac{\partial}{\partial t_{2k}}\nonumber,\\
Q_0^W&=\sum_{k\geq 1} \frac{B_{2k}u^{2(2k-1)}}{2k(2k-1)}\sum_{\substack{i,j\geq 0\\i+j=2k-2}} (-1)^{i+1}\frac{\partial^2}{\partial t_i\partial t_j}\nonumber.
\end{align}
Using Zassenhaus formula (cf. equation~\eqref{eqn:Zassenhaus}), we can decompose the operator $\exp(W)$ into the following form :
\begin{Prop} \label{prop:decompW}
\begin{equation*}
\exp(W)=\exp(\mathfrak{B}_t)\exp(\frac{1}{2}Q_t^W)\exp(P_t),
\end{equation*}
where $P_t$ is a first order differential operator defined in Lemma \ref{lemP}, and $Q_t^W$ is a second order differential operator defined in Lemma \ref{Q}.
\end{Prop}
This proposition follows from Lemma \ref{lemP} and Lemma \ref{Q} in Section 3.2. Using this proposition, we can transform equation (\ref{1}) into
\begin{equation}\label{relationint}
\exp(F_H(u,t))=\exp(\mathfrak{B}_t) \exp(\frac{1}{2}Q_t^W) \exp(P_t)\cdot\exp(F_K(t)).
\end{equation}
The operator $\exp(P)$ in equation (\ref{maineq}) is obtained immediately from $\exp(P_t)$ in Lemma \ref{lemP} after the change of variables $t_k=(2k-1)!!q_{2k+1}$ .

For $m>0$, let $L_m=X_m+Y_m,$ where
\begin{equation}
X_m=\sum_{k>0,k+m>0} (k+m)q_k\frac{\partial}{\partial q_{k+m}},\quad Y_m=\frac{1}{2}\sum_{a+b=m} ab\frac{\partial^2}{\partial q_a\partial q_b}.
\label{eqn:XYm>0}
\end{equation}
We then compare operators $\exp(\mathfrak{B}_t)$ and $\exp(\sum_{m>0} a_m X_m)$, where the coefficients $a_m$ are chosen in the way such that the following proposition holds:
\begin{Prop}\label{P4}
For any polynomial or power series $G$ in variables $u$ and $t_k$, $k\geq 0$, we have
$$\exp{(\sum_{m>0} a_mu^mX_m)}\cdot \left(\left.G\right\vert_{t_k=(2k-1)!!q_{2k+1}}\right)=\left.\left(e^{\mathfrak{B}_t} \cdot G\right)\right\vert_{t_k= \widetilde{\phi_k}(u,q)},$$
where the numbers $\{a_m\}$ are determined by equation~\eqref{eqn:a}, and $t_k= \widetilde{\phi_k}(u,q)$ is defined in Section 2.1 before.
\end{Prop}
We need to explain the notation `` $\cdot$ '' used in the above formula.
For a differential operator $\cal D$ and a function $f$, we always use the notation `` ${\cal D} \cdot f$ '' to represent the action of $\cal D$ on $f$, in order to distinguish from `` ${\cal D}f$ '' which could represent the multiplication of ${\cal D}$ and $f$ as operators when $f$ is considered as an operator acting on the space of functions by multiplication.

We will prove the above proposition using Lemma \ref{P5}, which is how we obtain the series on the right hand side of (\ref{eqn:a}) to represent the numbers $\{a_m\}$ in the first place.

Let
\begin{equation*}
Q_q^W=\left.Q_t^W\right\vert_{t_k = (2k-1)!!q_{2k+1}}.
\end{equation*}
Applying the change of variables  $t_k=\widetilde{\phi_k}(u,q)$ on both sides of equation (\ref{relationint}), and using Proposition \ref{P4} for $G=\exp(\frac{1}{2}Q_t^W) \exp(P_t)\cdot\exp(F_K(t))$, we can establish the following connection between $\exp(F_H(u,q))$ and $\exp(F_K(q))$:
\begin{Cor}\label{Cor:1stID}
$$\exp(F_H(u,q))=\exp(\sum_{m>0} a_mu^mX_m)\exp(\frac{1}{2}Q_q^W)\exp(P)\cdot\exp(F_K(q)).$$
\end{Cor}
This expression is very close to equation (\ref{maineq}) in Theorem \ref{main} now. To complete the proof
of Theorem~\ref{main}, we need to compare the following two operators:
\begin{equation*}
\exp(\sum_{m>0} a_mu^mX_m)\exp{(\frac{1}{2}Q_q^W)} \quad \mbox{and} \quad \exp(\sum_{m>0} a_mu^mL_m).
\end{equation*}
Since $L_m = X_m + Y_m$, we can use Zassenhaus formula to obtain the following decomposition
\begin{Prop}\label{Qvira}
$$\exp(\sum_{m>0} a_mu^mL_m)=\exp(\sum_{m>0} a_m u^m X_m) \exp(\frac{1}{2}Q^{+}),$$
where $Q^{+}$ is a second order differential operator defined in Corollary \ref{QplusL}
\end{Prop}
It turns out that the operators $Q^{+}$ and $Q_q^W$ are not the same. This fact explains in some sense
why it is difficult to compare operators $\exp(\sum_{m>0} a_mL_m) \exp(P)$ and $\exp(W)$ directly without
decomposing them first. The application of Zassenhaus formula in our proof singles out the major difference between these operators. Fortunately such difference does not affect Theorem~\ref{main} due to
the following proposition:

\begin{Prop} \label{Quadratic}
Let $Q^{+}_{odd}$ be the sum of terms in $Q^{+}$ involving only odd variables $q_{2k+1}$. Then,
$$Q_q^W=Q^{+}_{odd}.$$
\end{Prop}
We would like to remark that the expressions for $Q^{+}$ and $Q_q^W$ which arose during the process of applying
Zassenhaus formula are very different. The relation of these two operators as given in the above proposition is not obvious at all
when we first saw them.
To prove this proposition, we need to express these operators in simpler forms. Hence in Section 3.2, we introduce the formal power series $Q^B(x,y)$ in Lemma \ref{Q} and $Q(x,y)$ in equation (\ref{Qxy}) to represent $Q_t^W$ and $Q^{+}$ respectively. We then perform a transformation of $Q^{+}$ using the Gaussian integral and also use properties of Gamma function and beta functions to prove Proposition \ref{Quadratic} in Section 3.3.

Combining the above results, we have
\begin{align*}
&\exp(F_H(u,q))\\
=&\exp{(\sum_{m>0} a_mu^mX_m)} \exp(\frac{1}{2}Q^{+}_{odd})  \exp(P)\cdot \exp{(F_K(q))}\\
=&\exp{(\sum_{m>0} a_mu^mX_m)} \exp(\frac{1}{2}Q^{+}) \exp(P)\cdot\exp{(F_K(q))}\\
=&\exp{(\sum_{m>0} a_mu^mL_m)} \exp(P)\cdot\exp{(F_K(q))}.
\end{align*}
The first equality above follows from Corollary \ref{Cor:1stID} and Proposition \ref{Quadratic}.
The second equlity follows from the fact that $\exp{(F_K(q))}$ and the operator $\exp(P)$ only contain odd variables $q_{2k+1}$.
The third equality follows from Proposition \ref{Qvira}. This completes the proof of Theorem \ref{main}. As mentioned before, we can replace the operator $\exp(P)$ with the exponential of a linear combinations of Virasoro operators, using the fact that the Kontsevich-Witten tau-function satisfies the Virasoro constraints. This leads to a proof of Corollary 2, which will be given in Section 4.

In the rest part of this section, we will prove the above propositions.

\subsection{Proof Proposition \ref{P4}}
In this subsection, will prove Proposition \ref{P4} and also discuss properties of the sequence of numbers $a_{m}$ appeared in this
proposition. We first discuss change of variables using differential operators in the form $e^X$, where $X$ is a derivation. This is due to the following basic fact. Suppose we have two elements $f$ and $g$ from a commutative algebra, and $X$ is a derivation on the algebra satisfying Leibniz rule, that is, $X\cdot (fg)=(X\cdot f) g+ f(X\cdot g)$.  Then, it is easy to verify, by induction for instance, that
$$X^n\cdot(f g)=\sum_{i+j=n}(i+j)!\frac{X^i\cdot f}{i!}\frac{X^j\cdot g}{j!},$$
and
$$e^X\cdot (f g)= (e^X\cdot f)(e^X\cdot g).$$
It follows that applying an exponential of a derivation on a polynomial function or a series simply performs a change of variables. This works for series with infinite number of variables too.

In this paper, many problems involve differential operators and series with infinitely many variables. However, we will reduce such problems to those with just one variable and containing infinitely many terms. This method can simplify the computation process, and it has been frequently used in many papers, e.g. \cite{MK}. In other words, we will use power series in one variable to describe the action of the operators used in our context. For the knowledge of power series, including the compositional and multiplicative inverse of a power series, derivatives, composition of two power series and products, we refer the readers to the book \cite{GJ}.

First, we will consider $X_m$ and $\mathfrak{B}_t$, which are derivations appeared in the Virasoro operator $L_m$ and the operator $W$,
and are defined in equations \eqref{eqn:XYm>0} and \eqref{Bt} respectively. Since we need to deal with operators in the form $\exp(\sum_{m>0} a_m u^m L_m)$ and $\exp(W)$, it is necessary to discuss how the two operators
$$\exp(\sum_{m>0} a_mu^mX_m)\quad\mbox{and}\quad \exp(\mathfrak{B}_t)$$
behave as changes of variables. The connection between these operators is given by Proposition \ref{P4}.
 Since $X_m$ and $\mathfrak{B}_t$ are derivations,
  to prove Proposition \ref{P4}, we only need to consider the case $G(t)=t_n$, which has the following form
\begin{Lem}\label{P5}
There exists a unique sequence of numbers $\{a_m\}, m\geq 1$, such that, for all $n\geq 0$,
$$\exp{(\sum_{m>0} a_mu^mX_m)}\cdot q_{2n+1}=\left.\frac{1}{(2n-1)!!}\left(e^{\mathfrak{B}_t} \cdot t_n\right)\right\vert_{t_k= \widetilde{\phi_k}(u,q)}.$$
\end{Lem}
To prove this lemma, we will use operators with one variable $z$ to help us with the calculation. First we introduce an isomorphism $\Theta_1$ between the vector space of formal power series in variable $z$ and the vector space of linear functions in variables $q_i$:
\begin{align*}
\Theta_1: \{\sum_{i\geq 1}\alpha_iz^i  \left. \right\vert \alpha_i \mbox{ are constants} \} &\longrightarrow  \{ \sum_{i\geq 1} \alpha_iq_i  \left. \right\vert  \alpha_i \mbox{ are constants}\} \\
\Theta_1(\sum_{i\geq 1}\alpha_iz^i)  &= \sum_{i\geq 1} \alpha_iq_i.
\end{align*}
For a sequence of numbers $\{a_k\}$, let
$$\Phi_z^{+}=\sum_{k=1}^{\infty} a_kz^{1+k}\frac{\partial}{\partial z},
\quad \Phi_z^{-}=\sum_{k=1}^{\infty} a_k z^{1-k}\frac{\partial}{\partial z}. $$
The action of $e^{\Phi_z^{\pm}}$ on $z^n$ for $n\in\Z$ is of the form:
\begin{align}\label{phi}
&e^{\Phi_z^{\pm}}\cdot z^n \nonumber\\
=&z^n+\sum_{m>0} \frac{1}{m!}(\Phi_z^{\pm})^m\cdot z^n\nonumber\\
=&z^n+\sum_{m>0} \frac{1}{m!}\sum_{k_1,\dots k_m\geq 1} na_{k_1}(n\pm k_1)a_{k_2}\dots (n\pm \sum_{j=1}^{m-1} k_j)a_{k_m} z^{n\pm(\sum_{j=1}^{m} k_j)}.
\end{align}
The result of this action is a formal Laurent series. We can use this series to obtain the action of the following two operators on variables $q_n$ for $n\geq 1$ respectively:
$$\exp(\sum_{m<0} a_mX_m), \quad \exp(\sum_{m>0} a_mX_m).$$
For example, for the second case above, we have the following formula:
\begin{align*}
&\exp(\sum_{m>0} a_mX_m)\cdot q_n\\
=& q_n+\sum_{m>0} \frac{1}{m!}\sum_{k_1,\dots k_m\geq 1} na_{k_1}(n-k_1)a_{k_2}\dots (n-\sum_{j=1}^{m-1} k_j)a_{k_m}q_{n-(\sum_{j=1}^{m} k_j)},
\end{align*}
and this gives us a linear function in variables $q_i$. The coefficient of $q_i$ on the right hand side of the above equation is the same as the coefficient of $z^i$ with $i\geq 1$, in the series $\exp(\Phi_z^{-})\cdot z^n$. More precisely, let
\begin{equation}\label{Fz}
F(z)=e^{\Phi_z^{-}}\cdot z^n=z^n+\sum_{i=1}^{\infty} A_i^{(n)}z^{n-i},
\end{equation}
and define $(F(z))_{+}$ to be
$$(F(z))_{+}=z^n+\sum_{i=1}^{n-1} A_i^{(n)}z^{n-i}.$$
Then,
$$\Theta_1\left((e^{\Phi_z^{-}}\cdot z^n )_{+}\right)=\exp(\sum_{m>0} a_mX_m)\cdot q_n.$$
Note that in this formula, we can not replace $z^n$ by a formal power series $p(z) = \sum_{n=1}^{\infty} \alpha_n z^n$
since both sides of this equation are not well-defined in this case.
The action of $\exp(\sum_{m>0} a_mu^mX_m)$ on $q_n$ can be obtained by the variable change $q_k\rightarrow u^{-k}q_k$ in the above equation, that is
$$\exp(\sum_{m>0} a_mu^mX_m)\cdot q_n=q^n+\sum_{i=1}^{n-1} A_i^{(n)}u^iq_{n-i}.$$
Similarly, we can deduce that
\begin{equation}\label{Theta1plus}
\Theta_1\left(e^{\Phi_z^{+}}\cdot z^n \right)=\exp(\sum_{m<0} a_{-m}X_m)\cdot q_n.
\end{equation}
More generally, for any power series $p(z) = \sum_{n \geq 1} \alpha_n z^n$, we have
\begin{equation}
\Theta_1\left(e^{\Phi_z^{+}}\cdot p(z) \right)=\exp(\sum_{m<0} a_{-m} X_m)\cdot \Theta_1(p(z)).
\end{equation}
On the other hand, the action of $\exp(\mathfrak{B}_t)$ on $t_n$ is of the form:
\begin{align*}
&\exp(\mathfrak{B}_t)\cdot t_n\\
=& t_n+\sum_{m>0} \frac{1}{m!}\sum_{\substack{i+m=\sum_{j=1}^m 2k_j\\i,k_j\geq 1 }} \prod_{j=1}^m \frac{B_{2k_j}}{2k_j(2k_j-1)} u^{2i}t_{n-i}\\
=&\sum_{i=0}^{n}C_{i}u^{2i}t_{n-i},
\end{align*}
where the last step is implied by equation (\ref{B3}). Applying the variable change
$$t_k=\widetilde{\phi_k}(u,q)=\sum_{j=1}^{2k+1} \alpha_{j}^{(k)} u^{2k+1-j}q_j,$$
to the last equation, we have
\begin{align*}
\left.\left(e^{\mathfrak{B}_t}\cdot t_n\right)\right\vert_{t_k=\widetilde{\phi_k}(u,q)}\nonumber
=&\sum_{i=0}^{n}C_{i}u^{2i}\widetilde{\phi_{n-i}}(u,q)\nonumber\\
=&\sum_{i=0}^{n}C_{i}\sum_{j=1}^{2(n-i)+1} \alpha_j^{(n-i)}u^{2n+1-j} q_j.\\
%=&\sum_{j=1}^{2n+1} c_j^{(n)}u^{2n+1-j}q_{j},\nonumber
\end{align*}
Hence, in order to prove Lemma \ref{P5}, we only need to consider the case when $u=1$, that is,
\begin{equation}\label{withoutu}
\exp(\sum_{m>0} a_mX_m)\cdot q_{2n+1}=\frac{1}{(2n-1)!!}\sum_{i=0}^{n}C_{i}\widetilde{\phi_{n-i}}(1,q).
\end{equation}
The general case can also be obtained from the above equation after the change $q_k\rightarrow u^{-k}q_k$. Let
$$D=(1+z)^2z\frac{\partial}{\partial z},$$
which is the operator $\widehat{D}$ defined by equation (\ref{D}) with $u=1$. Under the isomorphism $\Theta_1$ we introduced before, the left hand side of equation (\ref{withoutu}) is equal to $\Theta_1\left((e^{\Phi_z^{-}}\cdot z^{2n+1})_{+}\right)$. And on the right hand side, by the definition of $\phi_k(u,z)$ and $\widetilde{\phi_{k}}(u,q)$ introduced in Section 2.1, we can see that
$$\Theta_1\left(D^{n-i}\cdot z \right)=\Theta_1\left(\phi_{n-i}(1,z)\right)=\widetilde{\phi_{n-i}}(1,q).$$
Therefore Lemma \ref{P5} follows from the next lemma:
\begin{Lem}\label{4}
There exists a unique sequence of numbers $\{a_m\}, m\geq 1$, for the operator $\Phi_z^{-}$, such that, for all $n\geq 0$,
\begin{equation}\label{6}
(e^{\Phi_z^{-}}\cdot z^{2n+1})_{+}=\frac{1}{(2n-1)!!}\sum_{i=0}^n C_iD^{n-i}\cdot z.
\end{equation}
\end{Lem}
{\em Proof}:
We first prove the uniqueness part by induction. Assume there exists a sequence $\{a_m\}, m\geq 1$, for operator $\Phi_z^{-}$, such that equation (\ref{6}) holds for all $n\geq 0$. When $n=0$, both sides of equation (\ref{6}) are equal to $z$. For $n=1$, we have, on the left hand side of equation (\ref{6}),
\begin{equation*}
(e^{\Phi_z^{-}}\cdot z^3)_{+}=z^3+3a_1z^2+(3a_1^2+3a_2)z.
\end{equation*}
On the right hand side, we have
$$Dz+C_1z=z^3+2z^2+\frac{13}{12}.$$
This gives us the only solution $a_1=2/3, a_2=-1/12$. Now we assume $\{a_i\}$, for $1\leq i \leq k,$ are all uniquely determined by equation (\ref{6}).
Then, we choose $n$ large enough such that $2n\geq k$. By equation (\ref{phi}), the coefficient of $z^{2n-k}$ in $(e^{\Phi_z^{-}}\cdot z^{2n+1})_{+}$ is
\begin{multline*}
A^{(2n+1)}_{k+1}=(2n+1)a_{k+1}+\\
\sum_{m\geq 2} \frac{1}{m!}\sum_{\substack{\sum_{j=1}^{m} k_j=k+1 \\ k_j\geq 1}} (2n+1)a_{k_1}\dots (2n+1- \sum_{j=1}^{m-1} k_j)a_{k_m} z^{2n+1-(\sum_{j=1}^{m} k_j)}.
\end{multline*}
The summation in the second term on the right hand side of the above equation consists of only $\{a_1,\dots,a_k\}$, and $A^{(2n+1)}_{k+1}$ is uniquely given by the right hand side of equation (\ref{6}). Hence there is only one solution for $a_{k+1}$. This proves the uniqueness.

Now we discuss the existence part. In order to find the coefficients $\{a_m\}$, we need to find a series $f$ such that $f = \exp(\Phi_z^{-})\cdot z$. To do so, we first define a function $f(z)$ to be
\begin{equation}\label{functionf}
f(z) = (-2\log(1-\frac{1}{1+z})-\frac{2}{1+z})^{-\frac{1}{2}}.
\end{equation}
Note that $f$ is a solution of the equation
\begin{equation}\label{Dff3}
Df=f^3.
\end{equation}
And its asymptotic expansion, which we also denote by $f$, can be computed in the following way (we explain why choosing this function in the remark later). By the equation above, we have
\begin{align*}
\frac{f^2}{z^2}&= \frac{-\frac{1}{2}z^{-2}}{\log(1-\frac{1}{1+z})+\frac{1}{1+z}}\\
 &= \frac{-\frac{1}{2}z^{-2}}{\log(\frac{1}{1+z^{-1}})+\frac{z^{-1}}{1+z^{-1}}}\\
 &= \frac{-\frac{1}{2}z^{-2}}{\sum_{n=1}^{\infty}(-1)^n\frac{z^{-n}}{n}-\sum_{n=1}^{\infty} (-1)^nz^{-n}}\\
 &= \frac{1}{1-2\sum_{n=3}^{\infty}(-1)^{n-1}\frac{n-1}{n}z^{2-n}}\\
 &= 1+2\sum_{n=3}^{\infty}(-1)^{n-1}\frac{n-1}{n}z^{2-n}+\dots .
\end{align*}
Hence the expansion of $f$ is of the form
$$f=z+\frac{2}{3}-\frac{1}{12}z^{-1}+\dots .$$
Note that the coefficient of $z$ in $f$ is set to be $1$, not $-1$. The above series $f$ determines a unique set of coefficients $\{a_m\}$ for $\Phi_z^{-}$. To see this, we set $f = \exp(\Phi_z^{-})\cdot z$,  then by equation (\ref{Fz}), for $n=1$ and $k\geq 1$,
$$f=\sum_{i=1}^{\infty}A_i^{(1)}z^{1-i},$$
where, by equation (\ref{phi}),
\begin{align*}
A^{(1)}_{1}&=a_1\nonumber;\\
A^{(1)}_{k}&=a_k+\sum_{m=2}^{k-1} \frac{1}{m!}\sum_{\substack{\sum_{j=1}^{m}k_j=k \\ k_j\geq 1} } a_{k_1}(1-k_1)a_{k_2}\dots (1-\sum_{j=1}^{m-1} k_j)a_{k_m}.
\end{align*}
Using induction on $k$, we can see that a fixed set $\{A^{(1)}_{k}\}$ uniquely determines $\{a_k\}$ and {\em vice versa}.
%We have computed the first four coefficients by hand, and they are $a_1=\frac{2}{3}, a_2=-\frac{1}{12}, a_3=\frac{7}{540}, a_4=-\frac{1}{1080}$.
From now on, we always assume $\{a_m\}$ are determined by
\begin{equation}\label{determinedam}
\exp(\Phi_z^{-})\cdot z=f,
\end{equation}
where $f$ is the function (\ref{functionf}). The existence part of Lemma \ref{4} follows from Lemma \ref{5} below. In other words, the coefficients $\{a_m\}$ determined here are exactly what we need for Lemma \ref{4}.
\begin{flushright}
$\Box$
\end{flushright}
{\bf Remark:} {\em (1)} The series $\exp(\Phi_z^{-})\cdot z^{-1}$ can be seen as the multiplicative inverse $f^{-1}$ of $f$. Since $\exp(\Phi_z^{-})\cdot 1=1$, we have
$$\exp(\Phi_z^{-})\cdot (z^{-1}z)=(\exp(\Phi_z^{-})\cdot z^{-1})f=1.$$
The standard definition of the multiplicative inverse of a power series usually requires that the series has a constant term (cf.\cite{GJ}). However, in our case, the series $f/z$ can be seen as a power series in $z^{-1}$ with the constant term $1$. Hence $f^{-1}$ can be understood as $z^{-1}(f/z)^{-1}$. From the discussion at the beginning of this section, we can deduce that, for $k\in\Z$,
$$f^k=e^{\Phi_z^{-}}\cdot z^k,$$
and for negative integer $k$, $f^k$ is a formal Laurent series.

\noindent
{\em (2)} To see where the function $f$ defined in equation (\ref{functionf}) comes from, we present a heuristic argument here. From the previous remark, we can see that equation (\ref{6}) can be transformed into
$$(f^{2n+1})_{+}=\frac{1}{(2n-1)!!}\sum_{i=0}^n C_iD^{n-i}\cdot z.$$
This gives us a hint that the series $f^{2n+1}$ behaves like the formal expression
$$\frac{1}{(2n-1)!!}\sum_{i=0}^{\infty} C_iD^{n-i}\cdot z.$$
This expression is not well-defined because it contains negative power of operator $D$. However, it gives us a possible equation that $f$ might satisfy. In particular, for $n=2$, since
$$\frac{1}{3}\sum_{i=0}^{\infty} C_iD^{2-i}\cdot z=\frac{1}{3}D\sum_{i=0}^{\infty} C_iD^{1-i}\cdot z,$$
we might expect $f$ to satisfy
$$f^5=\frac{1}{3}D\cdot f^3=f^2(D\cdot f),$$
which gives us
$$f^3=D\cdot f.$$
Solving this ODE we obtain a solution $f(z)$ which is the function defined by equation (\ref{functionf}).\begin{flushright}
$\Box$
\end{flushright}
\begin{Lem}\label{5}
For all $n\geq 0$,
$$(f^{2n+1})_{+}=\frac{1}{(2n-1)!!}\sum_{i=0}^n C_iD^{n-i}\cdot z.$$
\end{Lem}
{\em Proof}:
By the definition of function $f$ given in equation (\ref{functionf}), we have
$$e^{-\frac{1}{2}f^{-2}}=(1-\frac{1}{1+z})e^{\frac{1}{1+z}}.$$
Setting $(1+z)^{-1}=1-v$ and $f=x^{-1}$ in the above equation leads us to equation (\ref{vw}). Since equation (\ref{vw}) has two solutions $v(x)$ and $w(x)$, we obtain
$$(1+z)^{-1}=\pm \sum_{i=0}^{\infty}b_{2i+1}f^{-2i-1}-\sum_{i=1}^{\infty}b_{2i}f^{-2i}.$$
Now apply $D$ on both sides of the above equation. Since $Df=f^3$, we have $Df^{-i}=(-i)f^{-i+2}$. Hence
\begin{equation*}
z=\pm\sum_{i=0}^{\infty}(2i+1)b_{2i+1}f^{-2i+1}-\sum_{i=1}^{\infty}(2i)b_{2i}f^{-2i+2}.
\end{equation*}
It is easy to see that the highest power of $z$ in the series $f^k=\exp(\Phi_z^{-})\cdot z^k$ is $k$ for $k\in \Z$ by equation (\ref{phi}). Then $f^k$ for $k\leq 0$ will not contain terms of $z$ with positive degree. Since the coefficient of $z$ in $f$ is $1$, we can only take the positive sign in the above equation. This implies that
\begin{equation}\label{f}
f=z-\sum_{i=1}^{\infty}(2i+1)b_{2i+1}f^{-2i+1}+\sum_{i=1}^{\infty}(2i)b_{2i}f^{-2i+2}.
\end{equation}

From the equation $D\cdot f=f^3$, we have $D^n\cdot f=(2n-1)!!f^{2n+1}$. Then we can transform the equation in the lemma into
\begin{equation}\label{Dnf}
(D^n\cdot f)_{+}=\sum_{i=0}^n C_iD^{n-i} \cdot z.
\end{equation}
Furthermore, $D^n\cdot f^{-2i-1}$ will contain positive degree of $z$ if and only if $n>i$. Since $D^n\cdot f^{-2n}$ is a constant, $D^n\cdot f^{-2i}$ will not contain terms of $z$ with positive degree for all $n,i\geq 1$. Now we prove the lemma by induction.

When $n=0$, both sides of equation (\ref{Dnf}) are equal to $z$. When $n=1$, we apply $D$ on equation (\ref{f}) to obtain
\begin{align*}
(D\cdot f)_{+} &=(D\cdot z+(-3b_3)D\cdot f^{-1})_{+}\\
         &=D\cdot z+C_1z,
\end{align*}
where $C_1=3b_3$ by the relation (\ref{Cb}). So equation (\ref{Dnf}) holds for $n=1$. Assume equation (\ref{Dnf}) holds for all $n\leq k$. Let $n=k+1$. We apply $D^{k+1}$ on equation (\ref{f}) to obtain
\begin{align*}
(D^{k+1}\cdot f)_{+}&=D^{k+1}\cdot z+\sum_{i=1}^{k+1}(-1)(2i+1)b_{2i+1}(D^{k+1}\cdot f^{-2i+1})_{+}\\
              &=D^{k+1}\cdot z+\sum_{i=1}^{k+1}(-1)^{i+1}(2i+1)!!b_{2i+1}(D^{k+1-i}\cdot f)_{+} \\
              &=D^{k+1}\cdot z+\sum_{i=1}^{k+1}(-1)^{i+1}C_i\sum_{j=0}^{k+1-i}C_{k+1-i-j}D^j\cdot z\\
              &=D^{k+1}\cdot z+\sum_{j=0}^k(\sum_{i=1}^{k+1-j}(-1)^{i+1}C_iC_{k+1-j-i})D^j\cdot z \\
              &=D^{k+1}\cdot z+\sum_{j=0}^kC_{k+1-j}D^j\cdot  z,
\end{align*}
where the last step is implied by Lemma \ref{C}. Hence equation (\ref{Dnf}) holds for $n=k+1$. This completes the proof of the lemma and also completes the proof of Proposition \ref{P4}.
\begin{flushright}
$\Box$
\end{flushright}

\begin{Cor}\label{hz}
For the operator $\Phi_z^{+}$ with coefficients $\{a_m\}$ determined by equation (\ref{determinedam}), we define the series $h(z)$ to be
$$h(z)=e^{\Phi_z^{+}}\cdot z.$$
For the series $w(z)$ defined by equation (\ref{w}), we have
$$h(z)=\frac{1}{w(z)}-1=(\sum_{i=0}^{\infty} (-1)^{i}b_iz^i)^{-1} -1,$$
and
$$\frac{1}{h(z)}=\sum_{i=1}^{\infty}(-1)^{i-1}ib_iz^{i-2}.$$
In particular, since $w(z)$ is convergent and non-zero, $h(z)$ also converges for all $z \neq \infty$.
\end{Cor}
{\em Proof}:
We denote the compositional inverse function of $f$ to be $\psi(z)$. Since
\begin{equation*}
z=e^{-\Phi_z^{-}} e^{\Phi_z^{-}}\cdot z=e^{-\Phi_z^{-}}\cdot f(z)=f(e^{-\Phi_z^{-}}\cdot z),
\end{equation*}
we have
$$\psi=e^{-\Phi_z^{-}}\cdot z.$$
By equation (\ref{phi}), we have
\begin{align*}
h(z)=&e^{\Phi_z^{+}}\cdot z \\
=&z+\sum_{m>0} \frac{1}{m!}\sum_{k_1,\dots k_m\geq 1} (-1)(-a_{k_1})\dots (-1-\sum_{j=1}^{m-1} k_j)(-a_{k_m}) z^{1+(\sum_{j=1}^{m} k_j)}.
\end{align*}
On the other hand,
\begin{align*}
\psi^{-1}=&e^{-\Phi_z^{-}}\cdot z^{-1} \\
=&z^{-1}+\sum_{m>0} \frac{1}{m!}\sum_{k_1,\dots k_m\geq 1} (-1)(-a_{k_1})\dots (-1-\sum_{j=1}^{m-1} k_j)(-a_{k_m}) z^{-1-(\sum_{j=1}^{m} k_j)}.
\end{align*}
Compare the above two expressions, we obtain
$$h(z)=\psi^{-1}(z^{-1}) \quad\mbox{and}\quad\frac{1}{h(z)}=\psi(z^{-1}).$$
Since $\psi$ is the compositional inverse function of $f$, by equation (\ref{f}), we have
\begin{equation}\label{psi}
\psi=\sum_{i=1}^{\infty} (-1)^{i-1}ib_i z^{2-i}.
\end{equation}
Since
$$\mathrm{d}w/\mathrm{d}z=\sum_{i=1}^{\infty}(-1)^{i}ib_iz^{i-1},$$
we have
$$h(z)=z(\sum_{i=1}^{\infty}(-1)^{i-1}ib_iz^{i-1})^{-1}=-\frac{z}{\mathrm{d}w/\mathrm{d}z}.$$
By the relation $w^{'}(w-1)=zw$ from equation (\ref{vwrelation}), we obtain
\begin{equation}\label{handw}
h(z)=\frac{1}{w(z)}-1,
\end{equation}
which gives us the form of $h(z)$ used in this corollary and Theorem \ref{main}.
\begin{flushright}
$\Box$
\end{flushright}
Note that the coefficients $\{a_m\}$ can be computed easily using either the series $f(z)$ or $h(z)$. Also, since $h(z)=1/w(z)-1$, by equation (\ref{vw}), we have
\begin{equation}\label{formulaforh}
\frac{1}{1+h}e^{-\frac{1}{1+h}}=e^{-\frac{1}{2}z^2-1}.
\end{equation}

%Here we would like to mention that $h(z)$ is convergent for $z\geq 0$. To see this, we recall the {\bf Lambert W function}
%$$W(z)e^{W(z)}=z,$$
%with its series expansion
%$$W(z)=\sum_{n=1}^{\infty} \frac{(-n)^{n-1}}{n!}z^n.$$
%The radius of convergence for $W(z)$ is $e^{-1}$. Hence from equation (\ref{vw}) we can see that the series $w(z)$ is convergent at the domain $z\geq 0$ (in fact the radius of convergence is $\infty$). And since $0<w\leq 1$ and $h(z)=1/w(z)-1$, we deduce that $h(z)$ is convergent for $z\geq 0$. We will use this property in the next section.
\begin{Cor}\label{eta}
Let
$$\eta(u,z)=\exp(-\sum_{m=1}^{\infty} a_mu^mz^{1+m}\frac{\partial}{\partial z})\cdot z.$$
Then
\begin{equation}\label{invhz}
\eta(u,z)=\frac{1}{u}\sqrt{2\log(1+u z)-2+\frac{2}{1+u z}}.
\end{equation}
\end{Cor}
{\em Proof}:
Solving $z$ in terms of $h$ using equation (\ref{formulaforh}), we obtain
\[ z= \sqrt{2\log(1+h)-2+\frac{2}{1+h}}.\]
When $u=1$, the function $\eta(1,z)$ is the compositional inverse function of $h(z)$.
Therefore
$$\eta(1,z)=\sqrt{2\log(1+z)-2+\frac{2}{1+z}}.$$
Note that the coefficient of $z$ in $\eta(1,z)$ is set to be $1$. It is easy to see that
\begin{equation*}
u \eta(u,z)=\eta(1,u z).
\end{equation*}
This completes the proof.
\begin{flushright}
$\Box$
\end{flushright}

\subsection{Proof of Proposition \ref{prop:decompW} and Proposition \ref{Qvira}}
In this subsection, we will prove Proposition \ref{prop:decompW} and Proposition \ref{Qvira}. The main ingredient for
the proof of these propositions is the Zassenhaus formula, which can be considered as the inverse of
the  Baker-Campbell-Hausdorff (BCH) formula.
%Then, we will look for more relations between operators $\exp(\sum_{m>0}a_m u^m L_m)$ and $\exp(W)$.

The BCH formula is the expression of $Z$ as
$$Z=\log(e^{X}e^{Y})$$
for two elements $X$ and $Y$ from some Lie algebra over a field of characteristic zero. A general formula was introduced by Eugene Dynkin in \cite{ED}, expressing $Z$ as a sum of nested commutators involving $X$ and $Y$.
%\begin{equation}\label{BCH}
%Z=\sum_{n>0}\frac{(-1)^{n-1}}{n}
%\sum_{ \begin{smallmatrix} {r_i + s_i > 0} \\ {1\le i \le n} \end{smallmatrix}}
%\frac{[ X^{r_1} Y^{s_1} X^{r_2} Y^{s_2} \ldots X^{r_n} Y^{s_n} ]}{r_1!s_1!\cdots r_n!s_n!(\sum_{i=1}^n (r_i+s_i))},
%\end{equation}
%where $s_n$ and $r_n$ are non-negative integers, and
%$$[ X^{r_1} Y^{s_1} \ldots X^{r_n} Y^{s_n} ] = [ \underbrace{X,[X,\ldots[X}_{r_1} ,[ \underbrace{Y,[Y,\ldots[Y}_{s_1} ,\,\ldots\, [ \underbrace{X,[X,\ldots[X}_{r_n} ,[ \underbrace{Y,[Y,\ldots Y}_{s_n} ]]\ldots]].$$
%This term is always zero when $s_n>1$, or when $s_n=0$ and $r_n>1$.
If we add another parameter $\alpha$ to $Z$ by letting $Z=\log(e^{\alpha X}e^{\alpha Y})$, and write $Z=\sum_{m>0} \alpha^mZ_m$, the recursion relation in \cite{VSV} allows us to compute $Z_m$ starting from $Z_1=X+Y$. This relation and the general formula can be used in principal to construct the series $Z$ to arbitrary degree in terms of commutators. However, once we carry out the computation, we will meet its limitation very soon, since many identities involving the nested commutators, such as the Jacobi identity, are hidden inside the formula.

We use the notation `` $ad^j$ '' to denote the following nested commutator,
$$ad_X^0Y=Y, \quad ad_X Y=[X,Y], \quad ad_X^j Y=[X,ad_X^{j-1}Y].$$
The {\bf Zassenhaus formula} is known to be the inverse (or dual) of the BCH formula. It states that we can expand $e^{\alpha(X+Y)}$ as
\begin{equation}
e^{\alpha(X+Y)}=e^{\alpha X} e^{\alpha P_1}  e^{\alpha^2P_2} e^{\alpha^3P_3}\dots,
\label{eqn:Zassenhaus}
\end{equation}
where, $P_1=Y$, and for $n\geq 2$, $P_n$ is a homogeneous Lie polynomial in $X$ and $Y$ of degree $n$, (cf. \cite{WM}). The existence of such formula can be shown by using induction and the BCH formula on the product
\begin{equation}\label{Zinduction}
e^{-\alpha^n P_n} \dots e^{-\alpha P_1} e^{-\alpha X} e^{\alpha(X+Y)},
\end{equation}
for $n\geq 2$. In fact, the result in (cf. \cite{CMN}) shows that the term $P_{n}$ can be written as
\begin{equation}\label{Zassenhaus}
P_{n}=\frac{1}{n}\sum_{i=0}^{n-2} \frac{(-1)^{n-1}}{i!(n-1-i)!}ad_Y^iad_X^{n-1-i}Y.
\end{equation}
The Zassenhaus formula itself is a formal expression involving a product of infinite number of elements. When applying this formula on a specific case, one should check that the product of elements is well-defined and does not cause any problem of divergence. Later we will see that, in our case, the Zassenhaus formula works very well.

We mainly concern with the differential operators from the following three types:
$$\g_1=\left\{\left.X=\sum_{i\geq 0, j\geq 1}\alpha_{ij}t_{i}\frac{\partial}{\partial t_{i+j}} \right| \alpha_{ij} \mbox{ are constants}\right\},$$
$$\g_2=\left\{\left.Y=\sum_{i\geq 0}\beta_i\frac{\partial}{\partial t_i} \right| \beta_i\mbox{ are constants}\right\},$$
$$\g_3=\left\{\left.Z=\sum_{a,b\geq 0} \gamma_{ab}\frac{\partial^2}{\partial t_a\partial t_b}\right| \gamma_{ab}\mbox{ are constants, } \gamma_{ab}=\gamma_{ba}\right\}.$$
Each $\g_i$ is a Lie algebra, and so is the direct sum $g_1 \bigoplus g_2\bigoplus g_3$.  In particular, elements in $\g_2\bigoplus\g_3$ commute with each other. For $X,X^{'}\in \g_1, Y\in \g_2, Z\in \g_3$, we have
\begin{align*}
&[X,X^{'}] \in \g_1;\\
&[X,Y]= -\sum_{i\geq 0,j\geq 1} \alpha_{ij}\beta_i\frac{\partial}{\partial t_{i+j}} \in \g_2;\\
&[X,Z]= -\sum_{a,b\geq 0, j\geq 1} 2\alpha_{aj}\gamma_{ab}\frac{\partial^2}{\partial t_{a+j}\partial t_b} \in g_3.\\
\end{align*}
Suppose we apply the Zassenhaus formula on $\exp(X+Y)$. Then we can write the operator as
$$e^X e^{P^{(1)}_1} e^{P^{(1)}_2} e^{P^{(1)}_3} \dots,$$
where, by equation (\ref{Zassenhaus}),
$$P^{(1)}_n=\frac{(-1)^{n-1} }{n!}ad_X^{n-1}Y.$$
This is because $ad_X^{n-1}Y \in \g_2$ for all $n\geq 1$, which means that $ad_Y^iad_X^{n-1-i}Y=0$ for $i\geq 1$. Hence all operators $P^{(1)}_n$ commute with each other, and
\begin{equation}\label{Z1}
\exp(X+Y)=\exp(X) \exp(\sum_{n=1}^{\infty} P^{(1)}_n).
\end{equation}
Also, observe that $ad_X^{n-1}Y$ is a first order differential operator that only contains operators $\partial/\partial t_a$ with $a\geq n$. Therefore, in $\sum_{n=1}^{\infty} P^{(1)}_n$, the coefficient of $\partial/\partial t_a$ is always finite for all $a\geq 1$. Similarly, we can easily obtain
\begin{equation}\label{Z2}
\exp(X+Z)=\exp(X) \exp(\sum_{n=1}^{\infty} P^{(2)}_n)
\end{equation}
with
$$P^{(2)}_n=\frac{(-1)^{n-1}}{n!}ad_X^{n-1}Z,$$
where $P^{(2)}_n$ is a second order differential operator that only contains operators $\partial t_a\partial t_b$ with $a+b\geq n-1$. This guarantees that the operator $\sum_{n=1}^{\infty} P^{(2)}_n$ is well-defined. Our argument still works if we use variables $q_i$ with $i\geq 1$ for the operators. We can also check the validity of the two formulas (\ref{Z1}) and (\ref{Z2}) by applying the BCH formula on the right hand sides. And we refer the readers to the appendix for more details.

Now we apply Zassenhaus formula to some exponential operators used in our context.
We first prove Proposition~\ref{prop:decompW} which is the combination of Lemma~\ref{lemP} and Lemma~\ref{Q} below.
We will apply Zassenhaus formula to the decomposion
  \[ W=\mathfrak{B}_t+\frac{1}{2}Q_0^W+P_0 \]
   as given in equation (\ref{Ws}). It is easy to see that
 $\mathfrak{B}_t \in \g_1$, $P_0\in \g_2$, and $Q_0^W\in \g_3$.
\begin{Lem}\label{lemP}
\begin{equation}\label{P}
\exp(W)=\exp(\mathfrak{B}_t+\frac{1}{2}Q_0^W)\exp(P_t),
\end{equation}
where
$$P_t=-\sum_{i=1}^{\infty}C_{i}u^{2i}\frac{\partial}{\partial t_{i+1}}$$
with $C_i$ defined by equation (\ref{B3}).
\end{Lem}
{\em Proof}:
First we notice that $ad_{\mathfrak{B}_t}^{m-1}P_0 \in \g_2$ for all $m\geq 1$, and since $P_0$ commute with $Q_0^W$, we have
$$ad_{\mathfrak{B}_t+\frac{1}{2}Q_0^W}^{m-1}P_0=ad_{\mathfrak{B}_t}^{m-1}P_0.$$
From the discussion before, we can compute $P_t$ in equation (\ref{P}) as
$$P_t=\sum_{m=1}^{\infty} \frac{(-1)^{m-1}}{m!}ad_{\mathfrak{B}_t}^{m-1}P_0.$$
For, $m\geq 1$, the formula for the nested commutator $ad_{\mathfrak{B}_t}^{m-1}P_0$ is in fact
\begin{align}\label{formulaP}
&ad_{\mathfrak{B}_t}^{m-1}P_0 \nonumber\\
=&(-1)^m\sum_{i=m}^{\infty}\sum_{\substack{k_1,\dots,k_m\geq 1\\\sum_{j=1}^m k_j=i}}\left( \prod_{j=1}^m \frac{B_{2k_j}}{2k_j(2k_j-1)}\right)u^{4i-2m}\frac{\partial}{\partial t_{2i-m+1}}.
\end{align}
We can check this using induction. In order to simplify the expression of equations later, we define the number $\widetilde{B}_k$ to be
\begin{equation}\label{tildeB}
\widetilde{B}_k=\frac{B_{2k}}{2k(2k-1)}.
\end{equation}
When $m=1$,
$$ad_{\mathfrak{B}_t}^{0}P_0=P_0=-\sum_{i=1}^{\infty}\widetilde{B}_iu^{4i-2} \frac{\partial}{\partial t_{2i}}.$$
Assume the formula (\ref{formulaP}) holds for $m=n, n\geq 1$. Then, when $m=n+1$, we have
$$ad_{\mathfrak{B}_t}^{n}P_0=\left[\mathfrak{B}_t,ad_{\mathfrak{B}_t}^{n-1}P_0 \right].$$
For $k_{n+1}\geq 1$, since
$$\left[\sum_{j=0}^{\infty}t_j\frac{\partial}{\partial t_{j+2k_{n+1}-1}},\frac{\partial}{\partial t_{i}}  \right]\\
=-\frac{\partial}{\partial t_{i+2k_{n+1}-1}},$$
by the definition of $\mathfrak{B}_t$ in equation (\ref{Bt}), we have
\begin{align*}
&ad_{\mathfrak{B}_t}^{n}P_0\\
=&\left[\mathfrak{B}_t,ad_{\mathfrak{B}_t}^{n-1}P_0 \right]\\
=&(-1)^{n+1} \sum_{k_{n+1}=1}^{\infty} \widetilde{B}_{k_{n+1}} u^{2(2k_{n+1}-1)}\sum_{l=n}^{\infty}\sum_{\substack{k_1,\dots,k_{n}\geq 1\\\sum_{j=1}^{n} k_j=l}} \left( \prod_{j=1}^{n} \widetilde{B}_{k_j} \right) u^{4l-2n} \frac{\partial}{\partial t_{2l+2k_{n+1}-n}}\\
=&(-1)^{n+1}\sum_{i=n+1}^{\infty}\sum_{\substack{k_1,\dots,k_{n+1}\geq 1\\\sum_{j=1}^{n+1} k_j=i}}\left( \prod_{j=1}^{n+1} \widetilde{B}_{k_j} \right) u^{4i-2(n+1)}\frac{\partial}{\partial t_{2i-n}}.
\end{align*}
This proves the formula for $ad_{\mathfrak{B}_t}^{m-1}P_0$, and we obtain
$$P_t=-\sum_{i=1}^{\infty}\sum_{m=1}^{i}\frac{1}{m!}\sum_{\substack{k_1,\dots k_m\geq 1\\\sum_{j=1}^m 2k_j=i+m }}\left(\prod_{j=1}^m \widetilde{B}_{k_j} \right)u^{2i}\frac{\partial}{\partial t_{i+1}}.$$
Then the lemma follows from equation (\ref{B3}).
\begin{flushright}
$\Box$
\end{flushright}
Note that, by equation (\ref{Cb}), we can obtain the operator $P$ in Theorem \ref{main} from the change of variable as:
\begin{equation*}
P=\left.P_t\right\vert_{t_k = (2k-1)!!q_{2k+1}}.
\end{equation*}

Next, we define another linear map $\Theta_2$ from the space of power series of $x$ and $y$ to the space of second order differential operators with constant coefficients of variables $t_i$ by the following formula
$$\Theta_2(\sum_{i,j\geq 1}\alpha_{ij}x^iy^j)=\sum_{i,j\geq 1} \alpha_{ij}\frac{\partial^2}{\partial t_i\partial t_j},$$
where $\alpha_{ij}$ are constants. This map is an isomorphism when restricted to the subspaces
\begin{equation*}
\Theta_2: \{\sum_{i,j\geq 1}\alpha_{ij}x^iy^j \left. \right\vert \alpha_{ij}=\alpha_{ji} \} \longrightarrow  \{ \sum_{i,j\geq 1} \alpha_{ij}\frac{\partial^2}{\partial t_i\partial t_j}   \}
\end{equation*}
with the inverse map given by
\begin{equation*}
\Theta_2^{-1}( \frac{\partial^2}{\partial t_i\partial t_j})
= \frac{1}{2} (x^iy^j+x^jy^i).
\end{equation*}
Using this isomorphism, we have the following lemma.
\begin{Lem}\label{Q}
Define the series $Q^B(x,y)$ to be
\begin{align*}
Q^B(x,y)&=\frac{1-\exp(\mathfrak{B}(\frac{1}{x})+\mathfrak{B}(\frac{1}{y}))}{x+y}\\
&=\frac{1-\exp\left\lbrace \sum_{k=1}^{\infty}\frac{B_{2k}}{2k(2k-1)}(x^{2k-1}+y^{2k-1})\right\rbrace}{x+y}.\\
\end{align*}
Then
$$\exp(\mathfrak{B}_t+\frac{1}{2}Q_0^W)=\exp(\mathfrak{B}_t) \exp(\frac{1}{2}Q_t^W) $$
where
$$Q_t^W=\left.\Theta_2(Q^B(x,y))\right\vert_{t_k\rightarrow u^{-2k-1}t_k}.$$
%$$Q_t^W=\sum_{a,b\geq 0} Q^B_{ab}u^{2a+2b+2}\frac{\partial^2}{\partial t_a\partial t_b}.$$
\end{Lem}
{\em Proof}:
Using the Zassenhaus formula, we can see that
$$Q_t^W=\sum_{n=1}^{\infty} \frac{(-1)^{n-1}}{n!}ad_{\mathfrak{B}_t}^{n-1}Q_0^W.$$
Similar to the induction we have made in the proof of previous lemma, we can also derive the formula for the nested commutator $ad_{\mathfrak{B}_t}^{n-1}Q_0^W$. First, for simplicity, we will use the following convention
for summations and products:
 $\sum_l^k (\cdots) =0$ and $\prod_l^k (\cdots) =1$ for $l>k$. Using the notation (\ref{tildeB}), we have,
\begin{multline*}
ad_{\mathfrak{B}_t}^{n-1}Q_0^W=\sum_{k_1,\dots,k_n\geq 1}\left(\prod_{m=1}^{n} \widetilde{B}_{k_m}u^{2(2k_m-1)}\right)\\
\sum_{\substack{i,j\geq 0\\i+j=2k_1-2}}(-1)^{i+n} \sum_{l=0}^{n-1}\binom{n-1}{l}\frac{\partial^2}{\partial t_{i+\sum_{m=2}^{l+1} (2k_m-1)}\partial t_{j+\sum_{m=l+2}^n (2k_m-1)}}.
\end{multline*}
%\left(\frac{\partial^2}{\partial t_{i}\partial t_{j+\sum_{m=2}^n (2k_m-1)}}+\frac{\partial^2}{\partial t_{i+\sum_{m=2}^{n} (2k_m-1)}\partial t_{j}} \right.\\
%\left.
We prove this formula by induction on $n$.  In fact, when $n=1$,
\begin{align*}
ad_{\mathfrak{B}_t}^0Q_0^W=Q_0^W=\sum_{k\geq 1} \widetilde{B}_ku^{2(2k-1)} \sum_{\substack{i,j\geq 0\\i+j=2k-2}} (-1)^{i+1}\frac{\partial^2}{\partial t_i\partial t_j}.
\end{align*}
%When $n=2$,
%\begin{multline*}
%ad_{\mathfrak{B}_t}Q_0^W=\sum_{k_1,k_2\geq 1} \left(\prod_{m=1}^{2}\widetilde{B}_{k_m}u^{2(2k_m-1)}\right)\\ \sum_{\substack{i,j\geq 0\\i+j=2k_1-2}}(-1)^{i+2} \left(\frac{\partial^2}{\partial t_{i}\partial t_{j+2k_2-1}}+\frac{\partial^2}{\partial t_{i+2k_2-1}\partial t_{j}} \right)
%\end{multline*}
Assume that, for $n\geq 2$, the above formula for $ad_{\mathfrak{B}_t}^{n-1}Q_0^W$ holds. Since
$$\left[\sum_{j=0}^{\infty}t_j\frac{\partial}{\partial t_{j+2k_{n+1}-1}},\frac{\partial^2}{\partial t_{a}\partial t_b} \right]
=-\frac{\partial^2}{\partial t_{a}\partial t_{b+2k_{n+1}-1}}-\frac{\partial^2}{\partial t_{a+2k_{n+1}-1}\partial t_b},$$
we have
\begin{align*}
ad_{\mathfrak{B}_t}^{n}Q_0^W=&\left[\mathfrak{B}_t,ad_{\mathfrak{B}_t}^{n-1}Q_0^W \right]\\
=&\sum_{k_1,\dots,k_{n+1}\geq 1}\left(\prod_{m=1}^{n+1} \widetilde{B}_{k_m}u^{2(2k_m-1)}\right)\\
&\quad \sum_{\substack{i,j\geq 0\\i+j=2k_1-2}}(-1)^{i+n+1} \left\{\sum_{l=0}^{n-1}\binom{n-1}{l}\frac{\partial^2}{\partial t_{i+\sum_{m=2}^{l+1} (2k_m-1)}\partial t_{j+\sum_{m=l+2}^{n+1} (2k_m-1)}}\right. \\
&\quad\quad\quad  \left.+\sum_{l=1}^{n}\binom{n-1}{l-1}\frac{\partial^2}{\partial t_{i+\sum_{m=2}^{l+1} (2k_m-1)}\partial t_{j+\sum_{m=l+2}^{n+1} (2k_m-1)}} \right\}\\
=&\sum_{k_1,\dots,k_{n+1}\geq 1}\left(\prod_{m=1}^{n+1} \widetilde{B}_{k_m}u^{2(2k_m-1)}\right)\\
&\sum_{\substack{i,j\geq 0\\i+j=2k_1-2}}(-1)^{i+n+1} \sum_{l=0}^{n}\binom{n}{l}\frac{\partial^2}{\partial t_{i+\sum_{m=2}^{l+1} (2k_m-1)}\partial t_{j+\sum_{m=l+2}^{n+1} (2k_m-1)}}.
\end{align*}
%\frac{\partial^2}{\partial t_{i}\partial t_{j+\sum_{m=2}^{n+1} (2k_m-1)}}+\frac{\partial^2}{\partial t_{i+\sum_{m=2}^{n+1} (2k_m-1)}\partial t_{j}} \right.\\
%\left(\frac{\partial^2}{\partial t_{i}\partial t_{j+\sum_{m=2}^{n+1} (2k_m-1)}}+\frac{\partial^2}{\partial t_{i+\sum_{m=2}^{n+1} (2k_m-1)}\partial t_{j}} \right.\\
%&\left.+
Note that in the above equation, we have used the {\bf Pascal's rule}
$$\binom nl = \binom{n-1}{l-1} + \binom{n-1}l,$$
for $1\leq l\leq n-1$.
This shows that the above formula for the nested commutator $ad_{\mathfrak{B}_t}^{n-1}Q_0^W$ is valid.  Hence
\begin{multline*}
Q_t^W=\sum_{n=1}^{\infty}\frac{1}{n!}\sum_{k_1,\dots,k_n\geq 1}\left(\prod_{m=1}^{n} \widetilde{B}_{k_m}u^{2(2k_m-1)}\right)\\
\sum_{\substack{i,j\geq 0\\i+j=2k_1-2}}(-1)^{i+1} \sum_{l=0}^{n-1}\binom{n-1}{l}\frac{\partial^2}{\partial t_{i+\sum_{m=2}^{l+1} (2k_m-1)}\partial t_{j+\sum_{m=l+2}^n (2k_m-1)}}.
\end{multline*}
%Now, we show that the coefficient of $u^{2a+2b+2}\partial^2/\partial t_a\partial t_b$ in $Q_t^W$ is exactly $Q^B_{ab}$ from the power series $Q^B(x,y)$.
On the other hand, by the definition of $Q^B(x,y)$ in the lemma, we have
\begin{align*}
Q^B(x,y)=&-\sum_{n=1}^{\infty}\frac{1}{n!}\left( \mathfrak{B}(\frac{1}{x})+\mathfrak{B}(\frac{1}{y}) \right)^n\frac{1}{x+y}\\
=&-\frac{\left( \mathfrak{B}(\frac{1}{x})+\mathfrak{B}(\frac{1}{y}) \right)}{x+y}\sum_{n=1}^{\infty}\frac{1}{n!}\left( \mathfrak{B}(\frac{1}{x})+\mathfrak{B}(\frac{1}{y}) \right)^{n-1} . \\
\end{align*}
Since $x^{2k-1}+y^{2k-1}=(x+y)\sum_{i=0}^{2k-2}(-1)^{i}x^iy^{2k-2-i}$, we can expand $Q^B(x,y)$ as:
\begin{align*}
Q^B(x,y)=&\left(\sum_{k_1=1}^{\infty}\widetilde{B}_{k_1}\sum_{\substack{i,j\geq 0\\i+j=2k_1-2}}(-1)^{i+1}x^iy^j\right)
\sum_{n=1}^{\infty}\frac{1}{n!}\sum_{l=0}^{n-1}\binom{n-1}{l} \mathfrak{B}(\frac{1}{x})^l \mathfrak{B}(\frac{1}{y})^{n-1-l} \\
=&\left(\sum_{k_1=1}^{\infty}\widetilde{B}_{k_1}\sum_{\substack{i,j\geq 0\\i+j=2k_1-2}}(-1)^{i+1}x^iy^j\right)\\
&\quad\quad\sum_{n=1}^{\infty}\frac{1}{n!}\sum_{k_2,\dots,k_n\geq 1}\sum_{l=0}^{n-1}\binom{n-1}{l}
\left(\prod_{m=2}^{l+1}\widetilde{B}_{k_m}x^{2k_m-1}\right)\left(\prod_{m=l+2}^{n}\widetilde{B}_{k_m}y^{2k_m-1}\right) \\
&\quad\quad\\
&=\sum_{n=1}^{\infty}\frac{1}{n!}\sum_{k_1,\dots,k_n\geq 1}\left(\prod_{m=1}^{n} \widetilde{B}_{k_m}\right)\\
&\quad\sum_{\substack{i,j\geq 0\\i+j=2k_1-2}}(-1)^{i+1}\sum_{l=0}^{n-1}\binom{n-1}{l}x^{i+\sum_{m=2}^{l+1} (2k_m-1)}y^{j+\sum_{m=l+2}^n (2k_m-1)}.
\end{align*}
For the degree of $u$ in $Q_t^W$, observe that
$$i+\sum_{m=2}^{l+1} (2k_m-1)+j+\sum_{m=l+2}^n (2k_m-1)=-1+\sum_{m=1}^n (2k_m-1).$$
Then $Q_t^W$ can be obtained from $\Theta_2(Q^B(x,y))$ by the variable change $t_k\rightarrow u^{-2k-1} t_k$.
This completes the proof of the lemma and Proposition~\ref{prop:decompW}.
\begin{flushright}
$\Box$
\end{flushright}
Furthermore, we can write $Q^B(x,y)$ as the following expression, using the Stirling's approximation for the gamma function (\ref{gamma}):
\begin{equation}\label{GxGy}
Q^B(x,y)=\frac{1}{x+y}-\frac{x^{\frac{1}{x}}y^{\frac{1}{y}}e^{\frac{1}{x}+\frac{1}{y}}}{2\pi\sqrt{xy}}\Gamma(\frac{1}{x})\Gamma(\frac{1}{y})\frac{1}{x+y}.
\end{equation}
We will use the above expression later in our proofs. From the definition of $Q^B(x,y)$ in the previous lemma, we can see that its expansion is of the form
\begin{equation}\label{QBexpansion}
Q^B(x,y)= \sum_{i,j\geq 0} Q^B_{ij}x^iy^j,
\end{equation}
for some constants $Q^B_{ij}$ with  $Q^B_{ij}=Q^B_{ji}$. Now we check that the following action is well-defined:
$$\exp(\frac{1}{2}Q_t^W) \exp(P_t)\cdot\exp(F_K(t)).$$
The operator $\exp(P_t)$ performs a shift $t_k\rightarrow t_k-C_{k-1}u^{2k-2}$ for $k\geq 2$, which means that $t_0$ and $t_1$ remains unchanged, and other $t_k$ will be added by a term of $u$ with positive degree $2k-2$. Hence $\exp(P_t)\cdot\exp(F_K(t))$ is a well-defined formal series. Since the degree of $u$ in any term of the operator $Q_t^W$ is always positive, the coefficient of a fixed term in $\exp(\frac{1}{2}Q_t^W) \exp(P_t)\cdot\exp(F_K(t))$ can only be a finite summation. Hence the above action is well-defined.

We go back to equation (\ref{1}). By Lemma \ref{lemP}, Lemma \ref{Q} and the discussion above, we have concluded the relation (\ref{relationint}) presented at the beginning of this section.

\noindent
{\bf Remark:} We note that equation (\ref{relationint}) and the expression of $Q^B(x,y)$ are similar to results on page 11 of Givental's paper \cite{AG}. Givental's results also describe a relation between Kontsevich-Witten tau function and a generating function
for Hodge integrals with a parametrization which is different from that of the Hodge tau function considered in this paper. The power series which determines the second order differential operator in his paper was deduced from the theory of Frobenius manifolds. In our case, the series $Q^B(x,y)$ arises naturally from the Zassenhaus formula. It is interesting to observe that these two power series are essentially the same up to change of signs. \begin{flushright} $\Box$ \end{flushright}

To prove Proposition~\ref{Qvira}, we can use Zassenhaus formula to obtain a decomposition of the following form 
\begin{equation} \label{eqn:decompexpL}
\exp(\sum_{m>0} a_mu^mL_m)=\exp(\sum_{m>0} a_m u^m X_m) \exp(\frac{1}{2}Q^{+}).
\end{equation}
To compute the operator $Q^+$, we set
$$X^{+}=\sum_{m>0} a_mu^mX_m\quad \mbox{and} \quad Y^{+}=\sum_{m>0} a_m u^m Y_m,$$
where $X_m$ and $Y_m$ are defined in equation \eqref{eqn:XYm>0}.
Since $X^{+}\in \g_1, Y^{+}\in \g_3$, we have
\begin{equation}\label{Qplus}
\frac{1}{2}Q^{+}=\sum_{n=1}^{\infty}\frac{(-1)^{n-1}}{n!}ad_{X^{+}}^{n-1}Y^{+}.
\end{equation}
Such a decomposition is entirely similar to the case of $\exp(\sum_{m<0} a_{-m} u^{-m} L_m)$.  To see this, first, we define  $\g_1^{'}, \g_3^{'}$ to be
$$\g_1^{'}=\left\{\left.\sum_{i\geq 1, j\geq 1}\alpha_{ij}q_{i+j}\frac{\partial}{\partial q_i}\right|\alpha_{ij} \mbox{ are constants} \right\},$$
$$\g_3^{'}=\left\{\left.\sum_{a,b\geq 1} \gamma_{ab}q_aq_b \right| \gamma_{ab} \mbox{ are constants, }\gamma_{ab}=\gamma_{ba}\right\}.$$
By a similar discussion at the beginning of this subsection, we can see that $\g_1^{'}$,$\g_3^{'}$ and $\g_1^{'}\bigoplus\g_3^{'}$ are all Lie algebras.

Let $\Xi$ be the Lie algebra isomorphism from $\g_1^{'}\bigoplus\g_3^{'}$ to $\g_1\bigoplus\g_3$ (in variables $q_i$) defined by
$$\Xi(q_{i+j}\frac{\partial}{\partial q_i})=-\frac{i+j}{i}q_{i}\frac{\partial}{\partial q_{i+j}},$$
and
$$\Xi(q_iq_j)=-ij\frac{\partial^2}{\partial q_i\partial q_j}.$$
The fact that $\Xi$ is a Lie algebra isomorphism is due to Lemma \ref{Xiiso} in Appendix A.2. 

For $m<0$, we also write $L_m = X_m + Y_m$ where
$X_m$ is defined as in equation~\eqref{eqn:XYm>0}
and $$Y_m =\frac{1}{2}\sum_{a+b=-m} q_a q_b.$$
Set
$$X^{-}=\sum_{m<0} a_{-m}u^{-m}X_m\quad \mbox{and} \quad Y^{-}=\sum_{m<0} a_{-m}u^{-m}Y_m.$$
We can see that
$$\Xi(-X^{-})=X^{+} \quad \mbox{and} \quad \Xi(-Y^{-})=Y^{+}.$$
Also, by the Zassenhaus formula, we have
$$\exp(-\sum_{m<0} a_{-m}u^{-m}L_m)=\exp(-X^{-}) \exp(-\frac{1}{2}Q^{-}),$$
where
$$-\frac{1}{2}Q^{-}=\sum_{n=1}^{\infty}\frac{(-1)^{n-1}}{n!}ad_{-X^{-}}^{n-1}(-Y^{-}).$$
Since $\Xi$ preserves the commutator relation, we can easily deduce that
\begin{equation}\label{Qminus}
\Xi(-Q^{-})=Q^{+}.
\end{equation}
Inspired by Lemma 4.5 in \cite{MK}, we introduce a method to calculate the quadratic form $Q^{-}$ explicitly, and then obtain the form of $Q^{+}$ using the isomorphism $\Xi$.

Suppose $\{V_m\}$ is a set of differential operators indexed by negative integers satisfying
$$[V_m,V_n]=(m-n)V_{m+n}.$$
The first thing we want to do is to differentiate the operator $\exp(A(u))$ by $u$, where
$$A(u)=\sum_{m<0} \alpha_mu^{-m}V_m,$$
and $\alpha_m$ are constants. Then we have
\begin{Lem}\label{dm}
For a fixed sequence of numbers $\{\alpha_m\}$, there exists a unique sequence $\{d_m\}$, such that
$$\frac{\partial}{\partial u}e^{A(u)}=(\sum_{m<0}d_mu^{-m-1}V_m) e^{A(u)}.$$
\end{Lem}
Here we are dealing with differentiation of an operator-valued function. For more details about this and a proof of the above lemma, we refer the readers to Appendix A.3. Observe that, for negative integers $m$, the sets of operators $\{L_{m}\}$, $\{X_m\}$ and $\{z^{1-m}\partial/\partial z\}$ all satisfy the same commutator relation as $\{V_m\}$ defined before. Hence they all satisfy Lemma \ref{dm}.

Let
\begin{align*}
E(u)&=\exp(\sum_{m<0} \alpha_mu^{-m}L_m)\\
&=\exp(\sum_{m<0} \alpha_mu^{-m}X_m)\exp(\frac{1}{2}Q^{L}(u)),
\end{align*}
where $Q^{L}(u)$ is a quadratic form. Then,
\begin{align}\label{suspicious}
\frac{\partial}{\partial u}E(u)=&\left( \frac{\partial}{\partial u}\exp(\sum_{m<0} \alpha_mu^{-m}X_m)\right) \exp(\frac{1}{2}Q^{L}(u))\nonumber\\
\quad\quad & +\exp(\sum_{m<0} \alpha_mu^{-m}X_m) \left(\frac{\partial}{\partial u}\exp(\frac{1}{2}Q^{L}(u)) \right).
\end{align}
Consider the first term on the right hand side of equation (\ref{suspicious}). Since $\{X_m\}$ satisfy Lemma \ref{dm}, we have
$$\left( \frac{\partial}{\partial u}\exp(\sum_{m<0} \alpha_mu^{-m}X_m)\right) \exp(\frac{1}{2}Q^{L}(u))=\left(\sum_{m<0}d_mu^{-m-1}X_m \right)E(u).$$
For the second term on the right hand side of equation (\ref{suspicious}), its action on an arbitrary function $G$ in variables $q_i$ is
\begin{align*}
&\left\{\exp(\sum_{m<0} \alpha_mu^{-m}X_m) \left(\frac{\partial}{\partial u}\exp(\frac{1}{2}Q^{L}(u)) \right)\right\}\cdot G\\
=&\exp(\sum_{m<0} \alpha_mu^{-m}X_m)\cdot\left\{  \left(\frac{\partial}{\partial u}\exp(\frac{1}{2}Q^{L}(u)) \right) G\right\}\\
=&\exp(\sum_{m<0} \alpha_mu^{-m}X_m)\cdot \left\{ \left(\frac{\partial}{\partial u}\frac{1}{2}Q^{L}(u) \right) \left(\exp(\frac{1}{2}Q^{L}(u))G\right)\right\}\\
=&\left\{\exp(\sum_{m<0} \alpha_mu^{-m}X_m)\cdot \frac{\partial}{\partial u}\frac{1}{2}Q^{L}(u)\right\}\left\{\exp(\sum_{m<0} \alpha_mu^{-m}X_m)\cdot \left(\exp(\frac{1}{2}Q^{L}(u))G\right)\right\}\\
&=\left\{\exp(\sum_{m<0} \alpha_mu^{-m}X_m)\cdot \frac{\partial}{\partial u}\frac{1}{2}Q^{L}(u)\right\}E(u)\cdot G.
\end{align*}
Recall that we use `` $\cdot$ '' to represent the action of differential operators on functions. The third equality of the above equation follows from the fact that $X_m$ is a derivation. Since $G$ is arbitrary, we have the following equation for differential operators
\begin{align*}
&\exp(\sum_{m<0} \alpha_mu^{-m}X_m) \left(\frac{\partial}{\partial u}\exp(\frac{1}{2}Q^{L}(u)) \right)\\
=&\left\{\exp(\sum_{m<0} \alpha_mu^{-m}X_m)\cdot \frac{\partial}{\partial u}\frac{1}{2}Q^{L}(u)\right\}E(u).
\end{align*}
Hence,
$$\frac{\partial}{\partial u}E(u)=\left(\sum_{m<0}d_mu^{-m-1}X_m+\exp(\sum_{m<0} \alpha_mu^{-m}X_m)\cdot \frac{\partial}{\partial u}\frac{1}{2}Q^{L}(u)\right)E(u).$$

On the other hand, applying Lemma \ref{dm} on the case $V_m=L_m$, we have
$$\frac{\partial}{\partial u}E(u)=(\sum_{m<0}d_mu^{-m-1}L_m) E(u).$$
Therefore, since $E(u)$ is invertible,
\begin{equation}\label{Ym}
\frac{\partial}{\partial u}\frac{1}{2}Q^{L}(u)=\exp(-\sum_{m<0} \alpha_mu^{-m}X_m)\cdot\sum_{m<0}d_mu^{-m-1}Y_m.
\end{equation}
This gives us a way to compute $Q^{L}(u)$ when the coefficients $\{\alpha_m\}$, which also determines $\{d_m\}$ by Lemma \ref{dm}, are given. Note that by the definition of $Y_m$, $Y_{-1}$ is defined to be $0$. And we can see from the above equation that the power of $u$ for the term $q_iq_j$ in $Q^{L}(u)$ is always $i+j$. Usually, an easy way to obtain $\{d_m\}$ is to apply Lemma \ref{dm} on operators $\{z^{1-m}\partial/\partial z\}$, when we know the explicit form of the function $x(u,z)$ defined as
\begin{equation}\label{xz}
x(u,z)=\exp(\sum_{m<0}\alpha_{m}u^{-m}z^{1-m}\frac{\partial}{\partial z})\cdot z.
\end{equation}
Again, by Lemma \ref{dm},
\begin{align*}
&\frac{\partial}{\partial u}\exp(\sum_{m<0}\alpha_{m}u^{-m}z^{1-m}\frac{\partial}{\partial z})\\
=&(\sum_{m<0}d_{m}u^{-m-1}z^{1-m}\frac{\partial}{\partial z})\exp(\sum_{m<0}\alpha_{m}u^{-m}z^{1-m}\frac{\partial}{\partial z}).
\end{align*}
We use the operators on both sides of the above equation acting on $z$. Since
$$\frac{\partial}{\partial u}\exp(\sum_{m<0}\alpha_{m}u^{-m}z^{1-m}\frac{\partial}{\partial z})\cdot z = \frac{\partial}{\partial u}\left(\exp(\sum_{m<0}\alpha_{m}u^{-m}z^{1-m}\frac{\partial}{\partial z})\cdot z\right),$$
we have
\begin{equation}\label{xuz}
\frac{\partial}{\partial u}x(u,z)=(\sum_{m<0}d_{m}u^{-m-1}z^{1-m})\frac{\partial}{\partial z}x(u,z).
\end{equation}
So the coefficients $\{d_m\}$ can be easily obtained by comparing the partial derivatives of $x(u,z)$ with respect to $u$ and $z$.
Now, for the sequence of numbers $\{a_{-m}\}$ determined in Lemma \ref{4}, $m<0$, let $\alpha_m=-a_{-m}$, then $$Q^{L}(u)=-Q^{-}.$$
%Before we determine the quadratic form $Q^{-}$ using the above method, we recall some definitions related to the formal power series. The {\bf logarithmic series} $\log(1-x)$ is defined as
%$$\log(1-x)=-\sum_{i=1}^{\infty}\frac{x^i}{i!}.$$
%Suppose $p(x)$ is another formal power series, the composition $\log(1-p(x))$ is well-defined if and only if $p(x)$ has no constant term. Also, the derivative of $\log(1-x)$ is defined as
%$$\frac{\mathrm{d}}{\mathrm{d}x}\log(1-x)=-\frac{1}{1-x}=-\sum_{i=0}^{\infty}x^i.$$
%And
%\begin{equation}\label{Dxlog}
%\frac{\mathrm{d}}{\mathrm{d}x}\log(1-p(x))=-p^{'}(x)\frac{1}{1-p(x)}=-p^{'}(x)\sum_{n=0}^{\infty}p^n(x).
%\end{equation}
Now, to determine $Q^{-}$, it is convenient to consider the power series $Q(x,y)$ defined by the following function
\begin{equation}\label{Qxy}
Q(x,y) =\log\left((\frac{1}{h(x)}-\frac{1}{h(y)})(\frac{xy}{y-x})\right),
\end{equation}
where $h$ is the function defined in Corollary \ref{hz}.
%We assume that $x\neq y$ and $xy\neq 0$, so the series is in two independent variables and not trivial.
The series can be obtained in the following way. Using the series $1/h$ in Corollary \ref{hz}, we have
\begin{align*}
(\frac{1}{h(x)}-\frac{1}{h(y)})(\frac{xy}{y-x})&=xy\sum_{i=1}^{\infty}(-1)^{i-1}ib_i\frac{x^{i-2}-y^{i-2}}{y-x}\\
&=1+\sum_{i=3}^{\infty}(-1)^i ib_i\sum_{m=1}^{i-2}x^my^{i-1-m}.
\end{align*}
Since the series $\sum_{i=3}^{\infty}(-1)^i ib_i\sum_{m=1}^{i-2}x^my^{i-1-m}$ has no constant term, the logarithm of the above series is well-defined and we have
\begin{align*}
Q(x,y)&=\sum_{n=1}^{\infty}\frac{(-1)^{n-1}}{n}\left(\sum_{i=3}^{\infty}(-1)^i ib_i\sum_{m=1}^{i-2}x^my^{i-1-m}\right)^n.
\end{align*}
Let the series $Q(x,y)$ be of the form
\begin{equation}\label{Qxyexpansion}
Q(x,y)=\sum_{i,j\geq 1} Q_{ij}x^iy^j,
\end{equation}
for some constants $Q_{ij}$. From the definition (\ref{Qxy}) we can see that $Q(x,y)=Q(y,x)$. Hence $Q_{ij}=Q_{ji}$. Next, similar to the map $\Theta_2$, we define the linear map $\Theta_3$ from the space of power series of $x$ and $y$ to the space of quadratic functions with constant coefficients in variables $q_i$ by the following formula
$$\Theta_3(\sum_{i,j\geq 1}\alpha_{ij}x^iy^j )=\sum_{i,j\geq 1} \alpha_{ij}q_iq_j,$$
where $\alpha_{ij}$ are constants. This map is an isomorphism when restricted to the subspaces
\begin{align*}
&\Theta_3: \{\sum_{i,j\geq 1}\alpha_{ij}x^iy^j  \left. \right\vert \alpha_{ij}=\alpha_{ji} \} \longrightarrow  \{ \sum_{i,j\geq 1} \alpha_{ij}q_iq_j   \} \\
\end{align*}
with the inverse map given by
$$\Theta_3^{-1}(q_iq_j)= \frac{1}{2}(x^iy^j+x^jy^i).$$
Now we prove that
\begin{Lem}\label{LemmaQxy}
$$Q^{-}=\left.\Theta_3(Q(x,y))\right\vert_{q_k\rightarrow u^kq_k}.$$
\end{Lem}
{\em Proof}:
After setting $\alpha_m=-a_{-m}$, the function $x(u,z)$ defined in equation (\ref{xz}) becomes $\eta=\eta(u,z)$ defined in Corollary \ref{eta}. Since
$$\frac{\partial}{\partial u}\eta(u,z)=-\frac{\eta}{u}+\frac{z^2}{u(1+uz)^2}\frac{1}{\eta}$$
and
$$\frac{\partial}{\partial z}\eta(u,z)=\frac{z}{(1+uz)^2}\frac{1}{\eta},$$
we have
\begin{align*}
\frac{\partial}{\partial u}\eta(u,z)=&\left(\frac{z}{u}-\frac{(1+u z)^2}{u z}\eta^2\right)\frac{\partial}{\partial z}\eta(u,z)\\
=&\left(\sum_{n=2}^{\infty} (-1)^{n-1}\frac{4}{(n+1)n(n-1)}u^{n-2}z^n\right)\frac{\partial}{\partial z}\eta(u,z).
\end{align*}
From equation (\ref{xuz}), we can see that, in this case,
$$d_{-n+1}=(-1)^{n-1}\frac{4}{(n+1)n(n-1)} $$
for $n\geq 2.$
Let
$$Q^{-}=\sum_{i,j\geq 1} Q^{-}_{ij}u^{i+j}q_iq_j$$
for some constants $Q^{-}_{ij}$ with $Q^{-}_{ij}=Q^{-}_{ji}$.
Note that as explained in the remark after equation (\ref{Ym}), the power of $u$ before $q_i q_j$ should be
$i+j$.
Hence
$$-\frac{1}{2}\frac{\partial}{\partial u}Q^{-}=-\frac{1}{2}\sum_{i,j\geq 1} (i+j)Q^{-}_{ij}u^{i+j-1}q_iq_j.$$
On the other hand, by equation (\ref{Ym}),
\begin{equation}\label{Theta2minus}
-\frac{1}{2}\frac{\partial}{\partial u}Q^{-}=\exp(\sum_{m<0}a_{-m}u^{-m}X_m)\cdot \left(\sum_{n=2}^{\infty}d_{-n+1}u^{n-2}Y_{-n+1}\right).
%=&\exp(\sum_{m<0}a_{-m}u^{-m}X_m)\cdot  \left(\sum_{n=3}^{\infty}(-1)^{n-1}\frac{2}{(n+1)n(n-1)}u^{n-2}\sum_{i=1}^{n-2}q_iq_{n-1-i}\right).
\end{equation}
We want to find a series $T(x,y)$, such that
\begin{equation*}
T(x,y)=\sum_{n=3}^{\infty} \frac{1}{2}d_{-n+1}\sum_{i=1}^{n-2} x^iy^{n-1-i},
\end{equation*}
which gives us
$$\Theta_3(T(x,y))=\sum_{n=3}^{\infty}\frac{1}{2}d_{-n+1}\sum_{i=1}^{n-2}q_iq_{n-1-i}
= \sum_{n=2}^{\infty}d_{-n+1} Y_{-n+1}.$$
This series is nothing but the following one
$$T(x,y)=\frac{x}{x-y}(\frac{1+y}{y})^2\log(1+y)-\frac{y}{x-y}(\frac{1+x}{x})^2\log(1+x)-\frac{1}{x}-\frac{1}{y}-\frac{3}{2}.$$
One way to see this is that
\begin{align*}
&\frac{x-y}{xy}\sum_{n=3}^{\infty} \frac{1}{2}d_{-n+1}\sum_{i=1}^{n-2} x^iy^{n-1-i}\\
=&\sum_{n=3}^{\infty} (-1)^{n-1}\frac{2}{(n+1)n(n-1)}(x^{n-2}-y^{n-2})\\
=&\sum_{n=3}^{\infty}(-1)^{n-1}\left(\frac{1}{n+1}-\frac{2}{n}+\frac{1}{n-1}\right)(x^{n-2}-y^{n-2})\\
=&-\frac{1}{x^3}(\log(1+x)-x+\frac{x^2}{2}-\frac{x^3}{3})-\frac{2}{x^2}(\log(1+x)-x+\frac{x^2}{2})-\frac{1}{x}(\log(1+x)-x)\\
\quad &+\frac{1}{y^3}(\log(1+y)-y+\frac{y^2}{2}-\frac{y^3}{3})+\frac{2}{y^2}(\log(1+y)-y+\frac{y^2}{2})+\frac{1}{y}(\log(1+y)-y)\\
=&\frac{1}{y}(\frac{1+y}{y})^2\log(1+y)-\frac{1}{x}(\frac{1+x}{x})^2\log(1+x)-\frac{1}{y^2}+\frac{1}{x^2}-\frac{3}{2y}+\frac{3}{2x}\\
=&\frac{x-y}{xy}T(x,y).
\end{align*}
Note that $T(x,y)=T(y,x)$. Similar to the isomorphism $\Theta_1$  defined in Section 3.1, we define
$$\Theta_1(\sum_{i=1}^{\infty}\alpha_i x^i)=\Theta_1(\sum_{i=1}^{\infty}\alpha_i y^i)=\sum_{i=1}^{\infty}\alpha_i q_i$$
for constants $\alpha_i$. In fact, it is easy to see that, for two series $f_1(x), f_2(y)$
$$\Theta_3(f_1(x)f_2(y))=\Theta_1(f_1(x))\Theta_1(f_2(y)).$$

On the other hand,
since $ h^i(z) = e^{\Phi_z^+} \cdot z^i$ by the definition of $h$ in Corollary \ref{hz},
by equation (\ref{Theta1plus}), we have
$$\Theta_1(h^i(x))\Theta_1(h^j(y)) = \exp(\sum_{m<0}a_{-m}X_m)\cdot (q_iq_j) $$
for all $i, j \geq 1$.
This implies that
\begin{multline*}
\Theta_3\left(\sum_{n=1}^{\infty}\frac{1}{2}d_{-n+1}\sum_{i=1}^{n-2}h^i(x)h^{n-1-i}(y)\right)\\
=\exp(\sum_{m<0}a_{-m}X_m)\cdot  \left(\sum_{n=3}^{\infty}\frac{1}{2}d_{-n+1}\sum_{i=1}^{n-2}q_iq_{n-1-i}\right).
\end{multline*}
Then, by equation (\ref{Theta2minus}), and setting $u=1$, we have
$$\Theta_3\left(T(h(x),h(y))\right)=-\frac{1}{2}\sum_{i,j\geq 1} (i+j)Q^{-}_{ij}q_iq_j
= -\frac{1}{2} \Theta_3(\sum_{i,j\geq 1} (i+j)Q^{-}_{ij}x^iy^j).$$
The series $h$ has no constant term. Hence the composition $T(h(x),h(y))$ is well-defined. Since
 both sides of the above equation are symmetric with respect to $x$ and $y$,
we obtain
\begin{equation}\label{Txy}
T(h(x),h(y))=-\frac{1}{2}\sum_{i,j\geq 1}(i+j) Q^{-}_{ij}x^iy^j.
\end{equation}

To simplify the expression for $T(h(x), h(y))$, we take derivative of both sides of equation (\ref{formulaforh})
with respect to $z$ and replacing $e^{-\frac{1}{2} z^2 -1}$ by the left hand of the equation (\ref{formulaforh}). We then obtain the following formula for the derivative of $h(z)$:
$$\frac{\mathrm{d}h}{\mathrm{d}z}=\frac{z(1+h)^2}{h}.$$
Also, after taking the logarithm on equation (\ref{formulaforh}), we can obtain an expression for $\log(1+h)$,
$$\log(1+h)=\frac{1}{2}z^2+\frac{h}{1+h}.$$
Therefore,
\begin{equation*}
\frac{(1+h)^2}{h^2}\log(1+h)=\frac{1}{2}\frac{1}{h}z\frac{\mathrm{d}h}{\mathrm{d}z}+\frac{1}{h}+1.
\end{equation*}
Using the above equation, we obtain
\begin{align*}
T(h(x),h(y))&=\frac{h(x)}{h(x)-h(y)}\left(\frac{y}{2h(y)}\frac{\mathrm{d}h(y)}{\mathrm{d}y}+\frac{1}{h(y)}+1\right)\\
&\quad-\frac{h(y)}{h(x)-h(y)}\left(\frac{x}{2h(x)}\frac{\mathrm{d}h(x)}{\mathrm{d}x}+\frac{1}{h(x)}+1\right)-\frac{1}{h(x)}-\frac{1}{h(y)}-\frac{3}{2},\\
&=\frac{h(x)}{h(x)-h(y)}\frac{y}{2h(y)}\frac{\mathrm{d}h(y)}{\mathrm{d}y}-\frac{h(y)}{h(x)-h(y)}\frac{x}{2h(x)}\frac{\mathrm{d}h(x)}{\mathrm{d}x}-\frac{1}{2}
\\
&= \frac{h(x) h(y)}{2(h(x) - h(y))} \left( \frac{y}{h^2(y)} \frac{\mathrm{d}h(y)}{\mathrm{d}y}-\frac{1}{h(y)}
-\frac{x}{h^2(x)} \frac{\mathrm{d}h(x)}{\mathrm{d}x}+\frac{1}{h(x)} \right).
\end{align*}
By equation (\ref{Txy}), we have
\begin{align*}
\sum_{i,j\geq 1} Q^{-}_{ij}(i+j)x^iy^j&=(\frac{1}{h(x)}-\frac{1}{h(y)})^{-1}\left(\frac{\mathrm{d}(x/h(x))}{\mathrm{d}x}-\frac{\mathrm{d}(y/h(y))}{\mathrm{d}y}\right).
\end{align*}
We replace $x$ and $y$ by $ux$ and $uy$ respectively in the above equation. By the chain rule, we have
\begin{align}
\sum_{i,j\geq 1} Q^{-}_{ij}(i+j)u^{i+j-1}x^iy^j=(\frac{u}{h(u x)}-\frac{u}{h(u y)})^{-1}
\frac{\partial}{\partial u} \left(\frac{u}{h(u x)} - \frac{u}{h(u y)}\right). \label{pQ-}
\end{align}
By the formula of $1/h$ in Corollary \ref{hz}, we have
\begin{align*}
\frac{u}{h(u x)}-\frac{u}{h(u y)}
&=\frac{1}{x}-\frac{1}{y}+\sum_{n=3}^{\infty}(-1)^{n-1}nb_nu^{n-1}(x^{n-2}-y^{n-2})\\
 &=(\frac{1}{x}-\frac{1}{y})\left(1+\sum_{n=3}^{\infty}(-1)^{n}nb_nu^{n-1}\sum_{i=1}^{n-2}x^{i}y^{n-1-i}\right).
\end{align*}
Let
\[ K(x, y, u) = (\frac{1}{x}-\frac{1}{y})^{-1} \left(\frac{u}{h(u x)}-\frac{u}{h(u y)} \right)
= \frac{xy}{y-x} \left(\frac{u}{h(u x)}-\frac{u}{h(u y)}\right).
 \]
Then
\[ K(x, y, u) = 1+\sum_{n=3}^{\infty}(-1)^{n}nb_nu^{n-1}\sum_{i=1}^{n-2}x^{i}y^{n-1-i}
\]
and $\log K(x, y, u) $ is well-defined. Equation~\eqref{pQ-} can be written as
\[ \sum_{i,j\geq 1} Q^{-}_{ij}(i+j)u^{i+j-1}x^iy^j= K(x, y, u)^{-1} \frac{\partial}{\partial u} K(x, y, u)
= \frac{\partial}{\partial u} \log K(x, y, u).
\]
This shows that
both series
$\sum_{i,j\geq 1} Q^{-}_{ij}u^{i+j}x^iy^j$ and $\log K(x, y, u)$ have the same derivatives with respect to $u$
and both of them are $0$ at $u=0$.
Therefore we have
\begin{align*}
\sum_{i,j\geq 1} Q^{-}_{ij}u^{i+j}x^iy^j
&=\log K(x, y, u) =Q(ux,uy).
\end{align*}
Applying $\Theta_3$ on both sides of this equation, we obtain the lemma.
\begin{flushright}
$\Box$
\end{flushright}
{\bf Remark}: Until now, we have introduced two series $Q^B(x,y)$ and $Q(x,y)$. In this paper, we use the formal power series to describe the coefficients of our differential operators, and we only need to capture the values of the coefficients in the series. The values of $x$ and $y$ are not important to us.
\begin{flushright}
$\Box$
\end{flushright}

Using the relation (\ref{Qminus}), we can deduce that
\begin{Cor}\label{QplusL}
The operator $Q^+$ in equation \eqref{eqn:decompexpL} is given by
$$Q^{+}=\sum_{i,j\geq 1} Q_{ij}u^{i+j}ij\frac{\partial^2}{\partial q_i\partial q_j}.$$
\end{Cor}
This completes the proof of Proposition \ref{Qvira}.

\subsection{Proof of Proposition \ref{Quadratic}}
To prove this proposition, it is equivalent to show that
\begin{align*}
 Q^B_{ij}=(2i+1)!!(2j+1)!!Q_{2i+1,2j+1}.
\end{align*}
where $Q^B_{ij}$ are the coefficients of the series $Q^B(x,y)$ defined in equation (\ref{QBexpansion}) and $Q_{ij}$ are the coefficients of the series $Q(x,y)$ defined in equations (\ref{Qxy}) and (\ref{Qxyexpansion}), $i,j\geq 0$. Let
\begin{align*}
\Delta(x,y)&=x\frac{\partial}{\partial y}Q(x,y)\\
&=x\frac{\partial}{\partial y} \log\left((\frac{1}{h(x)}-\frac{1}{h(y)})(\frac{xy}{y-x})\right).
\end{align*}
Since $h=w^{-1}-1$ by equation (\ref{handw}), we have
\begin{equation}\label{twoterms}
\Delta(x,y)=x\frac{w^{'}(y)}{w(y)-w(x)}+\frac{x}{x-y}+\frac{xw^{'}(y)}{1-w(y)}+\frac{x}{y}.
\end{equation}
Using the Gaussian integral
$$\frac{1}{\sqrt{2\pi}}\int_{-\infty}^{\infty}z^{2k+2}e^{-\frac{z^2}{2}}\mathrm{d}z=(2k+1)!!, \quad k\geq 0,$$
We consider the following {\bf integral transform}
$$I(x,y)=\frac{1}{2\pi}\int\int_{\R^2} \Delta(xt,ys) e^{-\frac{t^2+s^2}{2}} \mathrm{d}t\mathrm{d}s.$$
On one hand, from equation (\ref{Qxyexpansion}), we can see that the expansion of $\Delta(x,y)$ is of the form
$$\Delta(x,y)=\sum_{i,j\geq 1} j Q_{ij} x^{i+1}y^{j-1}.$$
Hence the expansion of $I(x,y)$ is
\begin{equation}\label{Iexpansion}
I(x,y)=\frac{x}{y}\sum_{a,b\geq 0} Q_{2a+1,2b+1}(2a+1)!!(2b+1)!!x^{2a+1}y^{2b+1}.
\end{equation}
On the other hand, for the last two terms on the right hand side of equation (\ref{twoterms}),
$$\int_{\R}(\frac{xtw^{'}(ys)}{1-w(ys)}+\frac{xt}{ys})e^{-\frac{t^2}{2}}\mathrm{d}t=0 $$
since the integrant is skew symmetric with respect to $t$. Therefore
we only need to consider the integral transform of the first two terms of equation (\ref{twoterms}), namely
$$I(x,y)=\frac{1}{2\pi}(I_1+I_2),$$
where
\begin{align*}
I_1&=\int\int_{\R^2}xtw^{'}(ys)\frac{1}{w(ys)-w(xt)}e^{-\frac{t^2+s^2}{2}} \mathrm{d}t\mathrm{d}s, \\
I_2&=\int\int_{\R^2}\frac{xt}{xt-ys} e^{-\frac{t^2+s^2}{2}}\mathrm{d}t\mathrm{d}s.
\end{align*}
We first calculate the double integral $I_1$. Let $w_1=w(xt), w_2=w(ys)$. From now on we consider $w_1$ and $w_2$ as functions in variable $t$ and $s$ respectively, and treat $x$ and $y$ as non-zero constants. By the relation
$$x=w^{'}(x)(1-w^{-1}(x))$$
in equation (\ref{vwrelation}), we can transform $I_1$ into the following form
$$I_1=\int\int_{\R^2}\frac{1}{xy}\frac{\mathrm{d}w_1}{\mathrm{d}t}\frac{\mathrm{d}w_2}{\mathrm{d}s}(1-\frac{1}{w_1})\frac{1}{w_2-w_1}e^{-\frac{t^2+s^2}{2}} \mathrm{d}t\mathrm{d}s.$$
Observe that the two series $v$ and $w$ introduced in Section 2.3 have the relation $w(-z)=v(z)$. We extend the domain of $w$  to $\R$ by defining $w(z)=v(-z)$ for $z<0$. Then the range of $w$ is $(0,\infty)$. Also $w(xt)$ satisfies
$$w(xt)e^{1-w(xt)}=\left(e^{-\frac{t^2}{2}}\right)^{x^2}$$
by equation \eqref{vw}.
Hence
$$e^{-\frac{t^2}{2}}=w_1^{x^{-2}}e^{x^{-2}-w_1x^{-2}}.$$
The integral $I_1$ can be written as
$$I_1=\frac{e^{x^{-2}+y^{-2}}}{xy}\int_{0}^{\infty}\int_{0}^{\infty}(1-\frac{1}{w_1})\frac{1}{w_2-w_1}w_1^{x^{-2}}w_2^{y^{-2}}e^{-w_1x^{-2}-w_2y^{-2}} \mathrm{d}w_1\mathrm{d}w_2.$$
Consider the change of variables $(w_1,w_2)\rightarrow (z,u), w_1=x^2zu, w_2=y^2z(1-u)$. Since $w_1,w_2>0$, and
$$z=w_1x^{-2}+w_2y^{-2}, \quad u=\frac{w_1x^{-2}}{w_1x^{-2}+w_2y^{-2}},$$
we have $z\in (0,\infty)$ and $u\in (0,1)$. Then
$$I_1=\frac{e^{x^{-2}+y^{-2}}}{xy}(x^2)^{x^{-2}+1}(y^2)^{y^{-2}+1}\int_0^1 \int_0^{\infty}(1-\frac{1}{x^2zu})\frac{u^{x^{-2}}(1-u)^{y^{-2}}}{y^2-(x^2+y^2)u}z^{x^{-2}+y^{-2}}e^{-z}\mathrm{d}z\mathrm{d}u.$$
We solve the double integral in the above equation by two steps. First, we integrate by $z$. Using formula (\ref{euler2nd}), we have
\begin{align*}
&\frac{u^{x^{-2}}(1-u)^{y^{-2}}}{y^2-(x^2+y^2)u}\int_0^{\infty}(1-\frac{1}{x^2zu})z^{x^{-2}+y^{-2}}e^{-z}\mathrm{d}z\\
=&\frac{u^{x^{-2}-1}(1-u)^{y^{-2}}}{y^2-(x^2+y^2)u}\left(u\int_0^{\infty}z^{x^{-2}+y^{-2}}e^{-z}\mathrm{d}z- x^{-2}
    \int_0^{\infty} z^{x^{-2}+y^{-2}-1}e^{-z}\mathrm{d}z\right)\\
=&\frac{u^{x^{-2}-1}(1-u)^{y^{-2}}}{y^2-(x^2+y^2)u}\left(u\Gamma(x^{-2}+y^{-2}+1)-x^{-2}\Gamma(x^{-2}+y^{-2})\right)\\
=&\frac{(x^{-2}+y^{-2})u-x^{-2}}{y^2-(x^2+y^2)u}\Gamma(x^{-2}+y^{-2})u^{x^{-2}-1}(1-u)^{y^{-2}}\\
=&-x^{-2}y^{-2}\Gamma(x^{-2}+y^{-2})u^{x^{-2}-1}(1-u)^{y^{-2}}.
\end{align*}
Secondly, we integrate by $u$. Using formula (\ref{euler1st}), we have
$$\int_0^1u^{x^{-2}-1}(1-u)^{y^{-2}}\mathrm{d}u=B(x^{-2},y^{-2}+1).$$
Hence,
\begin{align*}
I_1&=-\frac{e^{x^{-2}+y^{-2}}}{xy}(x^2)^{x^{-2}}(y^2)^{y^{-2}}\Gamma(x^{-2}+y^{-2})B(x^{-2},y^{-2}+1)\\
&=-\frac{xe^{x^{-2}+y^{-2}}}{y}(x^2)^{x^{-2}}(y^2)^{y^{-2}}\frac{\Gamma(x^{-2})\Gamma(y^{-2})}{x^2+y^2},
\end{align*}
where the second step above is from equation (\ref{betaandgamma}). For the second integral $I_2$, we consider the change of variables $u=t-\frac{y}{x}s, z=s$. Then $I_2$ can be transformed into
\begin{align*}
I_2&=\int\int_{\R^2}\frac{u+\frac{y}{x}z}{u} e^{-\frac{(u+\frac{y}{x}z)^2}{2}}e^{-\frac{z^2}{2}}\mathrm{d}u\mathrm{d}z\\
&=\int\int_{\R^2}e^{-\frac{(u+\frac{y}{x}z)^2}{2}}e^{-\frac{z^2}{2}}\mathrm{d}u\mathrm{d}z+\frac{y}{x}\int\int_{\R^2}zu^{-1}e^{-\frac{(u+\frac{y}{x}z)^2}{2}}e^{-\frac{z^2}{2}}\mathrm{d}u\mathrm{d}z.\\
\end{align*}
Next, we will use the following two formulas of the Gaussian integral to calculate the above integral $I_2$:
For $a>0$,
$$\int_{-\infty}^{\infty} e^{-ax^2-bx}\,\mathrm{d}x=\sqrt{\frac{\pi}{a}}e^{\frac{b^2}{4a}} $$
and
$$\int_{-\infty}^{\infty} x e^{-ax^2-bx}\,\mathrm{d}x= -\frac{ \sqrt{\pi} b }{2}a^{-\frac{3}{2}}e^{\frac{b^2}{4a}}. $$

First, we compute
\begin{align*}
\int\int_{\R^2} e^{-\frac{(u+\frac{y}{x}z)^2}{2}} e^{-\frac{z^2}{2}} \mathrm{d}u \mathrm{d}z
=&\int_{-\infty}^{\infty} e^{-\frac{z^2}{2}(\frac{x^2+y^2}{x^2})} \int_{-\infty}^{\infty} e^{-\frac{1}{2}u^2-\frac{y}{x}zu} \mathrm{d}u\mathrm{d}z\\
=&\sqrt{2\pi} \int_{-\infty}^{\infty} e^{-\frac{z^2}{2}}\mathrm{d}z\\
=&2\pi,
\end{align*}
and
\begin{align*}
\int\int_{\R^2}zu^{-1}e^{-\frac{(u+\frac{y}{x}z)^2}{2}}e^{-\frac{z^2}{2}}\mathrm{d}u\mathrm{d}z
=&\int_{-\infty}^{\infty}u^{-1}e^{-\frac{u^2}{2}} \int_{-\infty}^{\infty} ze^{-\frac{x^2+y^2}{x^2}\frac{z^2}{2}-\frac{y}{x}uz}\mathrm{d}z\mathrm{d}u\\
=&-\sqrt{2\pi}\frac{y}{x}(\frac{x^2+y^2}{x^2})^{-\frac{3}{2}}\int_{-\infty}^{\infty}e^{-\frac{x^2}{x^2+y^2}\frac{u^2}{2}}\mathrm{d}u\\
=&-2\pi\frac{xy}{x^2+y^2}.
\end{align*}
Then we have
$$I_2=2\pi-2\pi\frac{y}{x}\frac{xy}{x^2+y^2}=2\pi\frac{x^2}{x^2+y^2}.$$
Finally, we obtain
\begin{align*}
\frac{y}{x}I(x,y)=&\frac{1}{2\pi}\frac{y}{x}(I_1+I_2)\\
=&\frac{xy}{x^2+y^2}-\frac{1}{2\pi}\frac{(x^2)^{x^{-2}}(y^2)^{y^{-2}}e^{x^{-2}+y^{-2}}}{x^2+y^2}\Gamma(\frac{1}{x^2})\Gamma(\frac{1}{y^2})\\
=& xyQ^B(x^2,y^2),
\end{align*}
where the last step above is by equation (\ref{GxGy}). Using equations (\ref{Iexpansion}) and (\ref{QBexpansion}), we can deduce that
$$(2a+1)!!(2b+1)!!Q_{2a+1,2b+1}=Q^B_{ab}.$$
This completes the proof of Proposition \ref{Quadratic} and thus also completes the proof of Theorem \ref{main}.
\begin{flushright}
$\Box$
\end{flushright}

\section{Proof of Corollary 2}
We prove Corollary \ref{em} in this section. Our following discussion is quite similar to the one in \cite{AE},
except that it is based on Theorem \ref{main} we have proved, and we also use the Zassenhaus formula,
which is more clear in terms of the computation here.
Later we can see that, since equation (\ref{hatb}) holds for any set of coefficients $\{l_m\}$,
the expression similar to equation (\ref{maineq}) that connects the two tau-functions using exponential
of Virasoro operators and first order differential operators is not unique.

The Virasoro constraints for the Kontsevich-Witten tau-function in variables $t$ are (cf. \cite{IZ}):
\begin{multline}
\widehat{L}_m=\sum_{k\geq m}\frac{(2k+1)!!}{(2k-2m-1)!!}t_{k-m}\frac{\partial}{\partial t_k}+\frac{1}{2}\sum_{k+l=m-1} (2k+1)!!(2l+1)!!\frac{\partial^2}{\partial t_k\partial t_l}\\
-(2m+3)!!\frac{\partial}{\partial t_{m+1}}+\frac{t_0^2}{2}\delta_{m,-1}+\frac{1}{8}\delta_{m,0},  (m=0,1,2,\dots)
\end{multline}
such that
$$\widehat{L}_m \cdot\exp(F_K(t)) =0,$$
for $m\geq -1$. If we remove the terms in $L_{2m}$ involving the even variables $q_{2k}$, then it coincides with the sum of the first two terms of the operator $\widehat{L}_m$ after the variable change $t_k=(2k-1)!!q_{2k+1}$. Now, we choose the Virasoro constraints $\widehat{L}_m$ with $m\geq 1$. Since $\exp(F_K(q))$ has no even variables $q_{2k}$, we can deduce that,
\begin{equation*}
\widetilde{V}_{2m}\cdot\exp(F_K(q))=0,
\end{equation*}
where
\begin{equation*}
\widetilde{V}_{2m}=L_{2m}-(2m+3)\frac{\partial}{\partial q_{2m+3}}.
\end{equation*}
Hence, for any sequence of numbers $\{l_{m}\}$, we have
\begin{equation*}
\exp(\sum_{m=1}^{\infty} l_{m}\widetilde{V}_{2m})\cdot\exp(F_K(q))=\exp(F_K(q)).
\end{equation*}
Using the Zassenhaus formula, we have
\begin{equation*}
\exp(\sum_{m=1}^{\infty} l_{m}\widetilde{V}_{2m})=\exp(\sum_{m=1}^{\infty} l_{m}L_{2m})\exp(-\sum_{k=1}^{\infty} \widetilde{b}_{k}\frac{\partial}{\partial q_{2k+3}}),
\end{equation*}
for a set of numbers $\{\widetilde{b}_{k}\}$ with
\begin{align}\label{bk}
\widetilde{b}_{1}&=5l_1;\nonumber\\
\widetilde{b}_{k}&=(2k+3)l_k+(2k+3)\sum_{n\geq 2}\frac{1}{n!} \sum_{\substack{m_j\geq 1\\\sum_{j=1}^nm_j=k}}l_{m_1}(3+2m_1)\dots l_{m_n}(3+2\sum_{j=1}^{n-1}m_j).
\end{align}
The above formula can be obtained using the methods introduced in the proof of Lemma \ref{lemP}. The induction is entirely similar. So here we will skip the details, and only give the formula for the nested commutator
\begin{align*}
&ad_{\sum_{m=1}^{\infty} l_{m}L_{2m}}^{n-1}(-\sum_{m=1}^{\infty}l_m(2m+3)\frac{\partial}{\partial q_{2m+3}})\\
=&(-1)^n\sum_{k=1}^{\infty}\sum_{\substack{m_j\geq 1 \\\sum_{j=1}^n m_j=k}}l_{m_1}(3+2m_1)\dots l_{m_n}(3+2\sum_{j=1}^{n}m_j) \frac{\partial}{\partial q_{2k+3}}.
\end{align*}
We can also see from equation (\ref{bk}) that $\{\widetilde{b}_{k}\}$ uniquely determine the coefficients $\{l_m\}$ and {\em vice versa}, using the same deduction in the proof of Lemma \ref{4}. And replacing $l_m$ with $u^{2m}l_m$, we have
$$\exp(\sum_{m=1}^{\infty} l_{m}u^{2m}\widetilde{V}_{2m})=\exp(\sum_{m=1}^{\infty} l_{m}u^{2m}L_{2m})\exp(-\sum_{k=1}^{\infty} \widetilde{b}_{k}u^{2k}\frac{\partial}{\partial q_{2k+3}}). $$
Hence, we can conclude that
\begin{equation}\label{hatb}
\exp(-\sum_{k=1}^{\infty} \widetilde{b}_{k}u^{2k}\frac{\partial}{\partial q_{2k+3}})\cdot\exp(F_K(q))=\exp(-\sum_{m=1}^{\infty} l_{m}u^{2m}L_{2m})\cdot\exp(F_K(q)).
\end{equation}
In our case, if we set $\widetilde{b}_{k}=b_{2k+1}$, then the operator $\exp(P)$ in Theorem \ref{main} can be replaced by the exponential of Virasoro operators at the right hand side of the above equation with the corresponding coefficients $\{l_m\}$. Another way to represent the relationship between the two sets of coefficients $\{b_{2k+1}\}$ and $\{l_m\}$, like we have done before, is to use a power series. Let
$$\theta(z)=\exp(-\sum_{m=1}^{\infty} l_{m}z^{1-2m}\frac{\partial}{\partial z})\cdot z.$$
Then, by equation (\ref{phi}),
\begin{align*}
\theta^{-3}(z)=& \exp(-\sum_{m=1}^{\infty} l_{m}z^{1-2m}\frac{\partial}{\partial z})\cdot z^{-3} \\
=& z^{-3}+3\sum_{n\geq 1} \frac{1}{n!}\sum_{m_1,\dots,m_n\geq 1}l_{m_1}(3+2m_1)\dots l_{m_n}(3+2\sum_{j=1}^{n-1}m_j)z^{-3-2\sum_{j=1}^n m_j}.\\
\end{align*}
And by equation (\ref{bk}), we have
$$\theta^{-3}(z)=3\sum_{k=0}^{\infty} \frac{b_{2k+1}}{2k+3}z^{-2k-3},$$
which gives us
$$\theta(z)=\left(3\sum_{k=0}^{\infty} \frac{b_{2k+1}}{2k+3}z^{-2k-3}\right)^{-\frac{1}{3}}.$$
Note that if we transform the above equation into
$$\frac{1}{3\theta^3(z)}-\frac{1}{3z^3}=\sum_{k=1}^{\infty} \frac{b_{2k+1}}{2k+3}z^{-2k-3},$$
this expression is very similar to equation (2.146) in \cite{AE}. The first few terms of $\theta(z)$ are
$$\theta(z)=z-\frac{1}{180}z^{-1}-\frac{67}{453600}z^{-3}+\dots$$
Furthermore, the following two operators can be combined into one using the Baker-Campbell-Hausdorff formula:
$$\exp(\sum_{m=1}^{\infty} a_mu^mL_m)\exp(-\sum_{m=1}^{\infty} l_{m}u^{2m}L_{2m})=\exp(\sum_{m=1}^{\infty} e_mu^mL_m).$$
Since $\{u^mL_m\}$ and $\{ z^{1-m}\partial/\partial z \}$ satisfy the same commutator relation as $V_m$, that is, $[V_m,V_n]=(m-n)V_{m+n}$, it is easy to see that
$$\exp(\sum_{m=1}^{\infty} a_{m}z^{1-m}\frac{\partial}{\partial z})\exp(-\sum_{m=1}^{\infty} l_{m}z^{1-2m}\frac{\partial}{\partial z})=\exp(\sum_{m=1}^{\infty} e_{m}z^{1-m}\frac{\partial}{\partial z}).$$
Using $\theta(z)$, we can obtain the coefficients $\{e_m\}$ from the following series
\begin{equation}\label{emtheta}
\exp(\sum_{m=1}^{\infty} e_{m}z^{1-m}\frac{\partial}{\partial z})\cdot z= \theta\left(\exp(\sum_{m=1}^{\infty} a_{m}z^{1-m}\frac{\partial}{\partial z})\cdot z\right)=\theta(f(z)),
\end{equation}
where $f(z)$ is defined in equation (\ref{functionf}). This proves Corollary \ref{em}. The first few terms of $\theta(f(z))$ are
$$\theta(f(z))=z+\frac{2}{3}-\frac{4}{45}z^{-1}+\frac{2}{45}z^{-2}+\dots. $$
The computation here is very straightforward, since both series $\theta$ and $f$ are given explicitly.
We have computed the first three coefficients $e_m$ by hand, they coincide with the data listed in
Alexandrov's paper \cite{A} (with the opposite sign, since our operators act on the Kontsevich-Witten tau-function,
not Hodge tau-function):
\begin{center}
  \begin{tabular}{ | l | l | l | l | }
    \hline
    $m$ & 1 & 2 & 3  \\ \hline
  $e_m$ & $\frac{2}{3}$ & $-\frac{2^2}{3^2\cdot 5}$ &  $\frac{2}{3^3\cdot 5}$ \\
    \hline
  \end{tabular}.
\end{center}

\section{Further remarks}
In the paper \cite{AE}, Alexandrov established the following relation between $\exp(F_H(u,q))$ and $\exp(F_K(q))$:
\begin{equation}\label{Alexconj}
\exp(F_K(q))=c(u) \widehat{G}_{+}\cdot\exp(F_H(u,q)),
\end{equation}
where $c(u)$ is an unknown Taylor series which is conjectured to be one, $$\widehat{G}_{+}=\beta^{-\frac{4}{3}L_0}\exp(\sum_{k=1}^{\infty}\widehat{a}_k\beta^{-k}L_k )\beta^{\frac{4}{3}L_0},$$
where the coefficients $\{\widehat{a}_k\}$ are determined by a series $f_{+}$, and $u=\beta^{1/3}$.
More precisely, $\{\widehat{a}_k\}$ are determined by the relation
$$f_{+}=\exp(\sum_{k=1}^{\infty} \widehat{a}_kz^{1-k}\frac{\partial}{\partial z})\cdot z $$
and $f_{+}$ is given as a solution of the following equation
$$\frac{f_{+}(z)}{1+f_{+}(z)}\exp(-\frac{f_{+}(z)}{1+f_{+}(z)})=E\exp(-E),$$
where
$$E=1+\sqrt{(\frac{1}{1+f_{+}(z)})^2+\frac{4}{3z^2}}.$$
The operator $\beta^{\frac{4}{3}L_0}$ is treated as the exponential of the degree operator $\frac{4}{3}\log(\beta)L_0$.
The action is
$$\exp(\frac{4}{3}\log(\beta)L_0)\cdot q_k=\beta^{\frac{4}{3}k}q_k.$$
The use of parameter $\beta$ in the conjectural equation (\ref{Alexconj}) is due to the relation between the Hodge and
Hurwitz tau-function \cite{MK}.
%
% In fact, it is very straightforward to check that equation (\ref{Alexconj}) is equivalent to
% $$\exp(F_K(q))=c(u) \exp(\sum_{k=1}^{\infty}\widehat{a}_ku^{k}L_k )\cdot\exp(F_H(u,q)).$$
% If Conjecture 2.1 in \cite{AE} is true, that is, $c(u)=1$. Then we expect to have the equality $\widehat{a}_k=-e_k$
% for all $k\geq 1$, where $\{e_k\}$ are the coefficients determined in equation (\ref{emtheta}).

At this moment it is not clear what is the relation between
$f_{+} $ and
\[ \exp(-\sum_{m=1}^{\infty} e_mz^{1-m}\frac{\partial}{\partial z})\cdot z .\]
Based on our known results, it is possible to derive an explicit equation that has $\theta(f(z))$ as its solution. But we believe that the formula (\ref{maineq}) may contain more information due to the appearance of numbers $C_k=(2k+1)!!b_{2k+1}$. If we switch to variables $t_i$ in $P$, the generating function after the shift
$$\exp(P)\cdot F_K(t)=F(t_0,t_1,t_2-C_1u^2,t_3-C_2u^4,\dots ),$$
which also satisfies the KdV hierarchy, looks particularly interesting. We think that there may exist some special connection between this shift and the one derived in \cite{MMBS}.

% The tau-functions and operators used in this paper are all in the form of multi-variable series. And our proofs are not
% in the language of Boson-Fermion correspondence. However, it would be very interesting to work entirely in the
% fermionic Fock spaces, and have the equivalent description of our results. For example, the way we treat the exponential
% of sums of differential operators, like Virasoro operators, is to first use the Zassenhaus formula to write it as a
% product of several ones, and then investigate the action of each one individually. There may be a connection between
% our methods and the description of Virasoro group action in \cite{AE}.

\section*{Appendix}
\subsection*{A.1 Special cases of Zassenhaus formula}
The Zassenhaus formula is not as well known as the BCH formula and it needs more caution when applying Zassenhaus formula.
In this appendix, we will verify formulas (\ref{Z1}) and (\ref{Z2}), which was deduced from Zassenhaus formula before, using the
following form of the BCH formula (cf.\cite{Miller}):
\begin{equation}\label{BCHintegral}
\log(e^Xe^Q)=X+Q-\left(\int_0^1\sum_{n=1}^{\infty}\frac{(1-e^{\text{ad}_X}e^{t \text{ad}_Q})^{n}}{n(n+1)}{\rm d}t\right)Q .
\end{equation}
Note that for the summation inside the integral above, $\text{ad}_X$ and $\text{ad}_Q$ are treated as two non-commutative  variables. Hence the integral will give us a formal power series in $\text{ad}_X$ and $\text{ad}_Q$.
Suppose $X\in \g_1$ and $Q\in \g_2(\mbox{or }\g_3)$. Then
$$e^{ \text{ad}_X}Q\in \g_2(\mbox{or }\g_3),$$
and, for any $Q^{'}\in \g_2(\mbox{or }\g_3)$,
$$e^{ t\text{ad}_Q}Q^{'}=Q^{'}.$$
This implies that, for $n\geq 0$,
$$(e^{\text{ad}_X}e^{t \text{ad}_Q})^nQ=(e^{\text{ad}_X})^nQ.$$
Hence, in this case, the formula (\ref{BCHintegral}) can be simplified into
\begin{align*}
\log(e^Xe^Q)&=X+Q-\left(\sum_{n=1}^{\infty}\frac{(1-e^{\text{ad}_X})^{n}}{n(n+1)}\right)Q\\
&=X+Q-\left(\sum_{n=1}^{\infty}\frac{(1-e^{\text{ad}_X})^{n}}{n}-\sum_{n=1}^{\infty}\frac{(1-e^{\text{ad}_X})^{n}}{n+1}\right)Q\\
&=X+Q-\left(-\log(e^{\text{ad}_X})+\frac{1}{1-e^{\text{ad}_X}}\log(e^{\text{ad}_X})+1\right)Q\\
&=X+Q-\left(\frac{e^{\text{ad}_X}}{1-e^{\text{ad}_X}}\text{ad}_X+1\right)Q\\
&=X-\left(\frac{e^{\text{ad}_X}}{1-e^{\text{ad}_X}}\text{ad}_X\right)Q\\
&=X+\left(\frac{\text{ad}_X}{1-e^{-\text{ad}_X}}\right)Q.
\end{align*}
Now we consider
\begin{align*}
Q&=\sum_{n=1}^{\infty}\frac{(-1)^{n-1}}{n!}ad_X^{n-1}Y\\
&=\left(\frac{1-e^{-\text{ad}_X}}{\text{ad}_X}\right)Y,
\end{align*}
for $Y\in \g_2(\mbox{or }\g_3).$ Then
\begin{align*}
\log(e^Xe^Q)&=X+\left(\frac{\text{ad}_X}{1-e^{-\text{ad}_X}}\frac{1-e^{-\text{ad}_X}}{\text{ad}_X}\right)Y\\
&=X+Y.
\end{align*}
This shows that the two formulas (\ref{Z1}) and (\ref{Z2}) are valid.

\subsection*{A.2 The Lie algebra isomorphism $\Xi$}
\begin{Lem}\label{Xiiso}
Let $\Xi$ be a linear map from $\g_1^{'}\bigoplus\g_3^{'}$ to $\g_1\bigoplus\g_3$ (in variables $q_i$) defined by
$$\Xi(q_{i+j}\frac{\partial}{\partial q_i})=-\frac{i+j}{i}q_{i}\frac{\partial}{\partial q_{i+j}},$$
and
$$\Xi(q_iq_j)=-ij\frac{\partial^2}{\partial q_i\partial q_j}.$$
Then $\Xi$ is a Lie algebra isomorphism.
\end{Lem}
{\em Proof}:
It is easy to see that the map $\Xi$ is a bijection by the definition. To check that $\Xi$ preserves the commutator relation, for example, we can check the following:

Since
$$[q_{i+j}\frac{\partial}{\partial q_i}, q_{i}\frac{\partial}{\partial q_{i-k}}]=q_{i+j}\frac{\partial}{\partial q_{i-k}};$$
$$[q_{i+j}\frac{\partial}{\partial q_i},q_iq_k]=q_{i+j}q_k, \mbox{for }i\neq k;$$
$$[q_{i+j}\frac{\partial}{\partial q_i},q_i^2]=2q_{i+j}q_i,$$
by the definition of $\Xi$,
\begin{align*}
[\Xi(q_{i+j}\frac{\partial}{\partial q_i}), \Xi(q_{i}\frac{\partial}{\partial q_{i-k}})]
=&[-\frac{i+j}{i}q_i\frac{\partial}{\partial q_{i+j}},-\frac{i}{i-k}q_{i-k}\frac{\partial}{\partial q_{i}}]\\
=&-\frac{i+j}{i-k}q_{i-k}\frac{\partial}{\partial q_{i+j}}\\
=&\Xi(q_{i+j}\frac{\partial}{\partial q_{i-k}}),
\end{align*}
and
\begin{align*}
[\Xi(q_{i+j}\frac{\partial}{\partial q_{i-k}}),\Xi(q_iq_k)]
=&[-\frac{i+j}{i}q_i\frac{\partial}{\partial q_{i+j}},-ik\frac{\partial^2}{\partial q_i\partial q_k}]\\
=&-(i+j)k\frac{\partial^2}{\partial q_{i+j}\partial q_k}\\
=&\Xi(q_{i+j}q_k);\\
[\Xi(q_{i+j}\frac{\partial}{\partial q_{i-k}}),\Xi(q_i^2)]
=&[-\frac{i+j}{i}q_i\frac{\partial}{\partial q_{i+j}},-i^2\frac{\partial^2}{\partial^2 q_i}]\\
=&-2(i+j)i\frac{\partial^2}{\partial q_{i+j}\partial q_i}\\
=&\Xi(2q_{i+j}q_i).
\end{align*}
The above calculations shows that $\Xi$ is an isomorphism.
\begin{flushright}
$\Box$
\end{flushright}

\subsection*{A.3 Differentiation of an operator-valued function}
The differentiation of an operator-valued function falls into the field of non-commutative calculus of operator derivatives. Derivatives of such functions have been discussed in the paper \cite{MS}, where operators considered are those acting on Banach spaces. Since we are not working with Banach spaces, we will not directly use formulas in \cite{MS}.  In our case, we treat $A=A(u)$ as a formal power series in $u$.
We denote the {\bf formal derivative} of $A(u)$ to be $A^{'}$ ,
$$A^{'}=\frac{\partial}{\partial u}A(u)=\sum_{m<0} (-m)\alpha_mu^{-m-1}V_m.$$
First we show that
$$\frac{\partial}{\partial u}e^{A}=\left(\sum_{n=1}^{\infty}\frac{1}{n!}ad_A^{n-1}A^{'}\right) e^{A}.$$
Similar results can be found in \cite{MS}. Since our setting is different from that in \cite{MS}, here we provide a proof of the above formula for completeness .

A straightforward computation shows that
\begin{align*}
\frac{\partial}{\partial u}e^{A}&=\sum_{n=1}^{\infty}\frac{1}{n!}\frac{\partial}{\partial u}A^n\nonumber\\
&=\sum_{n=1}^{\infty}\frac{1}{n!}\sum_{k=1}^{n}A^{k-1}A^{'}A^{n-k}\nonumber\\
&=\sum_{n=1}^{\infty}\frac{1}{n!} \left( nA^{n-1}A^{'}-\sum_{k=1}^{n}A^{k-1}ad_{A^{n-k}} A^{'}\right).
\end{align*}
Since
$ad_{A^2}A^{'}=2A(ad_AA^{'})-ad_A^2A^{'}$,
and $[A,ad_A]=0$, we can obtain, by induction,
$$ad_{A^k}A^{'}=\left(A^k-(A-ad_A)^k\right)A^{'}.$$
Using the following formula for the rising sum of binomial coefficients
$$\sum_{m=i}^{n}\binom{m}{i}=\binom{n+1}{i+1},$$
for any $i\geq 0, n\geq i$, we have
\begin{align*}
\frac{\partial}{\partial u}e^{A}&=\sum_{n=1}^{\infty}\frac{1}{n!}\sum_{k=1}^nA^{k-1}(A-ad_A)^{n-k}A^{'}\\
&=\sum_{n=1}^{\infty}\frac{1}{n!}\sum_{k=1}^n\sum_{i=0}^{n-k}A^{n-1-i}\binom{n-k}{i}(-1)^iad_A^i A^{'}\\
&=\sum_{n=1}^{\infty}\frac{1}{n!}\sum_{i=0}^{n-1}A^{n-1-i}\sum_{k=1}^{n-i}\binom{n-k}{i}(-1)^iad_A^i A^{'}\\
&=\sum_{n=1}^{\infty}\frac{1}{n!}\sum_{i=0}^{n-1}A^{n-1-i}\binom{n}{i+1}(-1)^iad_A^i A^{'}\\
&=\sum_{n=1}^{\infty}\sum_{i=0}^{n-1}\frac{1}{(n-1-i)!}A^{n-1-i}\frac{(-1)^i}{(i+1)!}ad_A^i A^{'}\\
&=\sum_{i=0}^{\infty}\sum_{n=i+1}^{\infty}\frac{1}{(n-1-i)!}A^{n-1-i}\frac{(-1)^i}{(i+1)!}ad_A^i A^{'}\\
&=e^A(\sum_{i=0}^{\infty} \frac{(-1)^{i}}{(i+1)!}ad_A^{i}A^{'}).
\end{align*}
Furthermore, using the formula
$$e^{ad_A} M=e^AMe^{-A},$$
and the equality
$$\sum_{\substack{i,j\geq 0\\i+j=m}}\frac{(-1)^j}{i!j!}=\frac{(1-1)^m}{m!}=0$$
for any $m\geq 1$, we have
\begin{align*}
\frac{\partial}{\partial u}e^{A}&=\left(\sum_{i,j=0}^{\infty}\frac{(-1)^j}{i!(j+1)!}ad_A^{i+j}A^{'}\right)e^{A}\\
 &=\left(\sum_{n=1}^{\infty}\frac{1}{n!}ad_A^{n-1}A^{'}\right) e^{A}.\\
\end{align*}
\begin{flushright}
$\Box$
\end{flushright}
{\bf Proof of Lemma \ref{dm}}:
Since $\{V_m\}$ satisfy the relation
$$[V_m,V_n]=(m-n)V_{m+n},$$
we can deduce that
$$\left(\sum_{n=1}^{\infty}\frac{1}{n!}ad_A^{n-1}A^{'}\right) e^{A}=(\sum_{m<0}d_mu^{-m-1}V_m) e^{A},$$
for some constants $\{d_m\}$. Note that $ad_A^{n-1}A^{'}$ consists of terms $u^{-m-1}V_m$ with $m\leq -n$. Hence $\sum_{m<0}d_mu^{-m-1}V_m$ is well-defined and $\{d_m\}$ are uniquely determined by $\{ \alpha_m \}$ and the bracket relation of
$\{ V_m \}$.
\begin{flushright}
$\Box$
\end{flushright}

\vspace{30pt} \noindent
Xiaobo Liu \\
Beijing International Center for Mathematical Research,\\
Peking University, Beijing, China.\\
E-mail address: {\it xbliu@math.pku.edu.cn}\\
\\
Gehao Wang \\
Beijing International Center for Mathematical Research,\\
Peking University, Beijing, China. \\
E-mail address: {\it gehao\_wang@hotmail.com}

\end{document}